\documentclass[pteplogo]{ptephy_v1}

\usepackage{amssymb}
\usepackage{amsmath}
\usepackage{bm}
\usepackage[hidelinks]{hyperref}
\usepackage{tikz}
\makeatletter
\newcommand{\Biggg}{\bBigg@{3.2}}
\makeatother

\usepackage{color}
\usepackage{ulem}

\newcounter{bla}

\newcommand{\TRENTo}{$\texttt{T{\raisebox{-0.15em}{\footnotesize R}}ENTo}$}
\DeclareMathOperator{\diag}{diag}

\begin{document}


\title{Efficient solver of relativistic hydrodynamics with implicit Runge-Kutta method\footnote{Report number: YITP-23-81, J-PARC-TH-0292}}

\author[1,2,4]{Nathan Touroux}
\affil{Department of Physics, Osaka University, Toyonaka, Osaka 560-0043, Japan\email{nathan.touroux@subatech.in2p3.fr}}
\author[1,2,3]{Masakiyo Kitazawa}
\affil{J-PARC Branch, KEK Theory Center, Institute of Particle and Nuclear Studies, KEK, 319-1106 Japan}
\author{Koichi Murase}
\affil{Yukawa Institute for Theoretical Physics, Kyoto University, Kyoto 606-8502 Japan}
\author{Marlene Nahrgang}
\affil{SUBATECH UMR 6457 (IMT Atlantique, Universit\'e de Nantes, IN2P3/CNRS), 4 rue Alfred Kastler, 44307 Nantes, France}

\begin{abstract}
We propose a new method to solve the relativistic hydrodynamic equations based
on implicit Runge-Kutta methods with a locally optimized fixed-point iterative
solver.  For numerical demonstration, we implement our idea for ideal
hydrodynamics using the one-stage Gauss-Legendre method as an implicit
method. The accuracy and computational cost of our new method are compared with
those of explicit ones for the (1+1)-dimensional Riemann problem, as well as
the (2+1)-dimensional Gubser flow and event-by-event initial conditions for
heavy-ion collisions generated by \TRENTo{}. We demonstrate that the solver
converges with only one iteration in most cases, and as a result, the implicit
method requires a smaller computational cost than the explicit one at the same
accuracy in these cases, while it may not converge with an unrealistically
large $\Delta t$. By showing a relationship between the one-stage
Gauss-Legendre method with the iterative solver and the two-step
Adams-Bashforth method, we argue that our method benefits from both the
stability of the former and the efficiency of the latter.
\end{abstract}

\subjectindex{D31, Quark-gluon plasma} 
\maketitle

\newpage

\section{Introduction}
Relativistic hydrodynamics is a versatile tool for describing various
long-range phenomena from astrophysics to nuclear physics at high energies. One
of the most important applications is relativistic heavy-ion collisions
(HIC). These collision experiments of heavy nuclei at ultrarelativistic
energies are currently operated at the Large Hadron Collider (LHC) at CERN, the
Relativistic Heavy Ion Collider (RHIC) at BNL, the Super Proton Synchrotron
(SPS) at CERN, and the Heavy Ion Synchrotron SIS18 at GSI, and several other
experiments are planned at large facilities all over the world. In the hot and
dense reaction zone, the strongly coupled quark-gluon plasma (QGP) is
created~\cite{Heinz:2000bk, Gyulassy:2004zy, Muller:2006ee}, which is well
described by a perfect fluid.  To investigate the properties of the QGP created
in heavy-ion collisions, we need to perform the simulation of heavy-ion
collision reactions based on the dynamical models and compare the results with
high-statistics experimental data~\cite{Teaney:2009qa, Jacak:2012dx,
Heinz:2013th, DerradideSouza:2015kpt, Braun-Munzinger:2015hba,
Romatschke:2017ejr, Busza:2018rrf}.  In modern modeling of heavy-ion
collisions, a relativistic viscous hydrodynamic code~\cite{Song:2007ux,
Chaudhuri:2008sj, Schenke:2010nt, Roy:2011pk, Bozek:2011ua, DelZanna:2013eua,
Karpenko:2013wva, Akamatsu:2013wyk, Murase:2015oie, Bazow:2016yra,
Okamoto:2016pbc, Pang:2018zzo, Du:2019obx} forms the core of the dynamical
description of the reactions together with an event-by-event initial-state
model and a final-state hadronic cascade.

To carry out those simulations, large computing resources and high performances
of numerical codes are required.  While in nature the evolution of a heavy-ion
collision takes several $10~\text{fm}$ from the initial state to the kinetic
freezeout of final particles, a numerical simulation of a single event can take
up to several hours depending on the computational devices. In addition, a
large number of collision events are required, particularly, for the ongoing
interest in global Bayesian analyses~\cite{Bayesian, BayesianNature,
Nijs:2020ors, Nijs:2020roc, Parkkila:2021tqq, Nijs:2021clz} and higher-order
fluctuation observables, such as the skewness and the kurtosis of the
event-by-event distributions of conserved charges~\cite{Asakawa:2015ybt,
Bluhm:2020mpc, STAR:2020tga, HADES:2020wpc}.  To properly extract the
properties of the QGP, both large computational resources and high-performance
codes are important.  In the present study, we focus on the latter: the
potential improvement of the numerical solver.

To efficiently solve relativistic hydrodynamics for heavy-ion collisions while
preserving the needed accuracy, various space-discretization schemes have been
implemented, such as SHASTA (sharp and smooth transport
algorithm)~\cite{Song:2007ux, Chaudhuri:2008sj, Roy:2011pk, Petersen:2008dd,
Holopainen:2010gz}, KT (Kurganov-Tadmor)~\cite{Schenke:2010nt, Bazow:2016yra,
Pang:2018zzo, Du:2019obx, McNelis:2021zji}, HLLE (Harten--Lax--van
Leer--Einfeldt)~\cite{Hirano:2000xa, DelZanna:2013eua, Karpenko:2013wva,
Tachibana:2014yai, Murase:2015oie}, and other Gudunov-type
schemes~\cite{Akamatsu:2013wyk, Okamoto:2016pbc}.  In this study, we instead
explore the time discretization.  The time integration methods are classified
into two, explicit and implicit methods. In an explicit method, the state of
the next time step is obtained by straightforward evaluations of given
expressions. In an implicit method, we need to solve implicit equations to
obtain the next state.  To the best of the authors' knowledge, the time
integration within all of the existing codes for HIC is performed by explicit
methods. The application of the implicit methods to the hydrodynamics in HIC is
an unexplored territory.  The goal of our work is to seek the possibility of
developing an implicit Runge-Kutta (RK) method that is stable, accurate, and
sufficiently efficient.

The computational cost of implicit methods is generally considered much higher
than explicit ones because implicit methods typically require numerically
solving equations, which would be one of the reasons that they have not been
applied in the hydrodynamics for HIC\@. However, it is only true for a
single-step time integration. There is another factor that affects the overall
computational cost to obtain the solution at the final time, i.e., the number
of time steps required to achieve sufficient accuracy and stability.  Here,
implicit methods have an important feature called the $A$-stability:
accumulated numerical errors are bounded for any positive time step, $\Delta
t$, for linear ordinary differential equations. One can, therefore, hope that
stable calculations can be performed using an implicit method at a larger
$\Delta t$ even for nonlinear equations leading to a smaller computational
cost.

Stability is also important in the implementation of computational
hydrodynamics.  Various technical procedures for stabilization, such as flux
limiters and monotonicity preservation, need to be developed using non-trivial
assumptions and careful tests for each.  Choosing a stable implicit method may
reduce such an effort for the stabilization, which would help deal with more
complex systems such as fluctuating hydrodynamics~\cite{Kapusta:2011gt,
Murase:2015oie, Murase:2016rhl, Singh:2018dpk, Bluhm:2018plm, Sakai:2020pjw},
non-equilibrium chiral fluid dynamics~\cite{Nahrgang:2011mg, Herold:2016uvv},
and magnetohydrodynamics~\cite{DelZanna:2013eua, Nakamura:2022idq}.  More
specifically, these complex systems are anticipated to be stiff systems, where
explicit methods require impractically small $\Delta t$ due to high-frequency
modes.  Genuine implicit methods can stably solve such stiff equation systems
with a moderate time step $\Delta t$.

Motivated by these advantages of implicit methods, we implement ideal
relativistic hydrodynamics with an implicit time integration and compare its
performance with a conventional explicit method quantitatively. To this end, we
employ the Kurganov-Tadmor scheme~\cite{KT} for the space discretization, which
is used for example in Refs.~\cite{Schenke:2010nt, Pang:2018zzo}. This spatial
scheme is independent of the time discretization and thus allows us to
systematically compare different time-discretization schemes. We will
especially use the one-stage Gauss--Legendre method as an implicit scheme and
Heun's method as an explicit scheme. Both are of second-order accuracy.

The most numerically demanding part of our implicit methods is the solver for
the Runge-Kutta nonlinear equations. In our code, we use a simple fixed-point
method for the solver but with several improvements. For example, we use the
solution found in the previous time step as the initial guess of the
fixed-point solver in the next step. We also check the convergence of
iterations cell by cell and skip updating the cells in which the iteration
converged.  We show that these optimizations enable an efficient implicit
scheme while preserving the accuracy and the simplicity of implementation.

We consider three initial conditions for benchmarks: the Riemann
problem~\cite{RiemannProblem}, the Gubser flow~\cite{Gubser}, and the
event-by-event \TRENTo{} initial conditions for heavy-ion
collisions~\cite{Trento}. We show that contrary to the typical expectations,
our implicit method is usually more efficient and accurate than the explicit
one even with the same $\Delta t$ for the cases of these initial conditions.
We also find that the fixed-point iteration may fail to converge with
large $\Delta t$, where the Courant-Friedrichs-Lewy (CFL)
condition~\cite{Courant1928upd, Courant1967pde} is unsatisfied. Nevertheless,
as far as a solution is obtained, its stability is ensured.

We argue that the better performance of our method comes from the fact that it
can be interpreted as the combination of the one-stage Gauss-Legendre (GL1) and
the two-step Adams-Bashforth (AB2) methods. We show that our method with a
single iteration can be related to AB2, which has the accuracy order $p=2$ but
less stability, while it converges to GL1 after iterations. Multiple iterations
are needed only when AB2 diverges from GL1. In this sense, our method makes use
of GL1 and AB2 depending on the local stiffness, thereby exploiting the
advantages in stability and efficiency of both methods in an optimal
way. Because of this property, our method is expected to be advantageous even
when applied to more complex systems of equations such as viscous
hydrodynamics.

This paper is organized as follows. In Sec.~\ref{sec:hydro}, we review the
hydrodynamic equations and their analytical solutions that will be used in
later sections. In Sec.~\ref{sec:discretization}, we next review the implicit
and explicit RK methods and the spatial discretization scheme in the present
study with technical details. In Sec.~\ref{sec:time}, we propose new
improvements to implicit RK methods and discuss the accuracy, efficiency, and
stability of the improved implicit method. The numerical setup and analyses are
given in Sec.~\ref{sec:setup}. We then discuss the numerical results in
Sec.~\ref{sec:results}. The final section is devoted to discussions and a
summary.

\section{Relativistic hydrodynamic}
\label{sec:hydro}
In this section, we summarize hydrodynamic equations and a few known analytical
solutions that will be used in Sec.~\ref{sec:results} for numerical tests.

\subsection{Hydrodynamics equations}

Hydrodynamic equations are given by the conservation law. We focus on the
conservation of the energy--momentum tensor $T^{\mu\nu}$:
\begin{align}
    \partial_{;\,\mu} T^{\mu\nu}
    &= \partial_{\mu}T^{\mu\nu}
    +\Gamma^{\mu}_{~\alpha\mu}T^{\alpha\nu}
    +\Gamma^{\nu}_{~\alpha\mu}T^{\mu\alpha} \nonumber \\
    &= 0\,, 
    \label{tmunucons}
\end{align}
where $\partial_{;\,\mu}$ denotes the covariant derivative,
and $\Gamma^{\nu}_{~\mu\alpha}$ the associated Christoffel symbols.
For the general case including further conserved currents, the scheme described
in this paper can be naturally extended.

In this paper we consider the ideal hydrodynamics where the energy--momentum
tensor is written in terms of thermodynamic quantities as
\begin{equation}
    T^{\mu\nu} = \epsilon u^\mu u^\nu - P\Delta^{\mu\nu}\,,
    \label{constitutive}
\end{equation}
where $u^\mu$ is the four-velocity with $u^\mu u_\mu = 1$, $\epsilon$ the
energy density, $P$ the pressure, and $\Delta_{\mu\nu} = g_{\mu\nu} - u_\mu
u_\nu$ the projector orthogonal to $u^\mu$. Equation~\eqref{constitutive}
contains five undetermined variables $(u^i,\epsilon,P)$ with $i=x,y,z$. Since
Eq.~\eqref{tmunucons} has four equations, one needs another constraint to close
the equations. This constraint is provided by the equation of state (EOS),
which is the pressure as a function of the energy density
$P(\epsilon)$. Equations~\eqref{tmunucons} and~\eqref{constitutive} together
with the EOS define the ideal hydrodynamics.

In the Cartesian coordinates, we choose the \textit{conserved
densities}\footnote{In this paper, for convenience, we mean by the
\textit{conserved densities} the densities of conserved charges even though the
densities in each cell are not conserved separately.}, $T^{tt}$, $T^{tx}$,
$T^{ty}$, and $T^{tz}$, as the state variables to be solved numerically. This
is a requirement from the space discretization scheme as we will see later. In
this treatment, we need an explicit way to calculate $T^{i\mu}$ as a function
of $T^{t\mu}$ that satisfies Eq.~\eqref{constitutive}. We first obtain
$\epsilon$ and $u^\mu$ from $T^{t\nu}$ and then calculate $T^{i\nu}$ using
Eq.~\eqref{constitutive}. To numerically determine $\epsilon$ and $u^\mu$
satisfying Eq.~\eqref{constitutive}, it is convenient to solve the following
equations~\cite{Karpenko:2013wva} for the energy density $\epsilon\;(>0)$ and
the velocity $v\;(\in[0,1])$ iteratively:
\begin{align}
    \epsilon &= T^{tt} - v K, &
    v &= \frac{K}{T^{tt}+P(\epsilon)},
    \label{eq:determine-ev}
\end{align}
where $K = \sqrt{\sum_{i=1}^3 (T^{ti})^2}$. These equations require the condition $T^{tt} \ge K$, which 
is ensured by the prescription described in Sec.~\ref{sec:space}. Then, $u^\mu$ is given by
\begin{align}
    u^i &= \frac{T^{ti}}{\sqrt{(T^{tt}+P(\epsilon))(\epsilon+P(\epsilon))}}, &
    u^t &= \sqrt{1+(u^i)^2}.
    \label{u0i}
\end{align}

In high-energy heavy-ion collisions, the space-time evolution of the produced
medium is well described by the Bjorken assumption of longitudinal
boost-invariance~\cite{bjorken}. In this case, it is convenient to employ the
Milne coordinates
\begin{align}
    \tau &= \sqrt{t^2-z^2}, &
    \eta &= \tanh^{-1}{\frac{z}{t}},
\end{align}
or written inversely,
\begin{align}
    t &= \tau\cosh{\eta}, & z &= \tau\sinh{\eta},
\end{align}
where $z$ is the \textit{longitudinal} direction parallel to the beam axis
while the \textit{transverse} coordinates, $x$ and $y$, remain unchanged.  In
this coordinate system, the metric is $g_{\mu\nu} = \diag(1,-1,-1,-\tau^2)$,
and the non-zero Christoffel symbols are $\Gamma^{\eta}_{~\tau\eta} =
\Gamma^{\eta}_{~\eta\tau} = 1/\tau$ and $\Gamma^{\tau}_{~\eta\eta} = \tau$.

To describe the dynamics in the mid-rapidity region of a high-energy collision,
it is also useful to reduce the spatial dimension to two by assuming the
\textit{boost invariance} in the Milne coordinate system where $\partial_\eta =
0$,\footnote{Here, when we apply $\partial_\eta$ to tensors, we assume that the
local basis of the tangent space is chosen to be the natural basis of the Milne
coordinate system. The relation $\partial_\eta = 0$ does not hold in mixed
basis cases such as $\partial_\eta u^t$ and $\partial_\eta u^z$.} $u^\eta =
0$,\footnote{More specifically, the flow velocity can be written in the Milne
coordinates as $(u^\tau, u^x, u^y, u^\eta) = (u^\tau(\tau,x,y), u^x(\tau,x,y),
u^y(\tau,x,y), 0)$. This corresponds to $(u^t,u^x,u^y,u^z) =
(u^\tau(\tau,x,y)\cosh\eta, u^x(\tau,x,y), u^y(\tau,x,y),
u^\tau(\tau,x,y)\sinh\eta)$ in the Cartesian coordinate basis.} and
$T^{\tau\eta} = 0$. In this case, the conservation equations~(\ref{tmunucons})
are given by
\begin{equation}
    \partial_\tau T^{\tau\nu} = -\partial_x T^{x\nu} -\partial_y T^{y\nu} - \frac{[\epsilon+P(\epsilon)]u^\tau u^\nu}{\tau},\label{dTMilne} \quad (\nu=\tau,x,y). \nonumber
\end{equation}
It is practically useful to rewrite Eq.~\eqref{dTMilne} as
\begin{align}
    \partial_\tau\tilde{T}^{\tau\tau} &= -\partial_x \tilde{T}^{x\tau} -\partial_y \tilde{T}^{y\tau} - P(\epsilon), \label{eq:milnet} \\
    \partial_\tau\tilde{T}^{\tau l} &= -\partial_x \tilde{T}^{xl} -\partial_y \tilde{T}^{yl}, \qquad (l=x,y),\label{eq:milnel}
\end{align}
with $\tilde{T}^{\mu\nu} = \tau T^{\mu\nu}$ and to treat $\tilde{T}^{\mu\nu}$ as the state variables.

\subsection{Analytical solutions}
In this subsection, we show two known analytical solutions of the hydrodynamic
equations, to which we will compare our numerical results in later sections.

\subsubsection{Riemann problem}
\label{sec:Riemann}

Let us consider a system having translational invariance in the $y$ and $z$
directions in the Cartesian coordinates.  We consider a Riemann problem
\cite{RiemannProblem} in the remaining $x$ direction given by the initial
condition, \newcommand{\epla}{\epsilon_{\text{plateau}}}
\newcommand{\emin}{\epsilon_{\text{min}}}
\newcommand{\emax}{\epsilon_{\text{max}}}
\begin{align}
    &\epsilon(t=0, x) = \begin{cases}
        \emax, & x<0,\\
        \emin, & x\geq 0,
    \end{cases}
    \label{Riemann:init}\\
    &u^t(t=0,x) = 1,
    \qquad
    u^x(t=0,x) = 0,
    \label{Riemann:init2}
\end{align}
with a discontinuity at $x=0$.  The analytical solution is given for a
particular choice of an EOS, $P(\epsilon)=c_s^2 \epsilon$. Here, $c_s$ is the
speed of sound. For $t>0$, the solution $\epsilon(x,t) =
\Tilde{\epsilon}(\Tilde{x})$ with $\tilde{x} = x/t$ is given
piecewise~\cite{RiemannProblem}:
\begin{align}
    \Tilde{\epsilon}(\Tilde{x}) &= \begin{cases}
       \epsilon_\text{max}, & \Tilde{x} < -c_s,\\
       \epsilon_\text{max}\left(\frac{(1-c_s)(1-\Tilde{x})}{(1+c_s)(1+\Tilde{x})}\right)^{\frac{1+c_s^2}{c_s}}, & -c_s \leq \Tilde{x}<
 -v_\text{r-p}, \\
       \epsilon_\text{plateau}, & -v_\text{r-p}\leq \Tilde{x}<v_\text{shock},\\
       \epsilon_\text{min}, & v_\text{shock} < \Tilde{x},
    \end{cases}\\
    u^x(\tilde x) &= \frac{v(\tilde\epsilon(\tilde x))}{\sqrt{1-v(\tilde\epsilon(\tilde x))^2}},
    \qquad
    u^t(\tilde x) = \sqrt{1+(u^x(\tilde x))^2},
\end{align}
where
\begin{align}
    v_\text{r-p} &= \frac{v(\epla) - c_s}{1-v(\epla) c_s}, \\
    v(\epsilon) &= \frac{1-(\epsilon/\epsilon_\text{max})^{2 c_s/(1+c_s^2)}}{1+(\epsilon/\epsilon_\text{max})^{2 c_s/(1+c_s^2)}}.
\end{align}
The constants $\epsilon_\text{plateau}$ and $v_\text{shock} = -v_r$ are
obtained by solving the following equations for $(\epla, v_l, v_r)$:

\begin{align}
    \epla\gamma(v_l)v_l &= \emin\gamma(v_r)v_r, \label{momentumcons}\\
    4\epla\gamma(v_l)v_l^2+\epla &= \emin\gamma(v_r)v_r^2+\emin, \label{Taub}\\
    v_l &= \frac{v(\epla)+v_r}{1+v(\epla)v_r},
\end{align}
where $\gamma(v) = 1/\sqrt{1-v^2}$.  We note that Eqs.~\eqref{momentumcons}
and~\eqref{Taub} represent the Taub equations~\cite{Taub:1948zz,
Landau:1987fm}.  The solution $\tilde\epsilon(\tilde x)$ is illustrated in
Fig.~\ref{fig:riemannsol}. The discontinuity at $\tilde x = v_\text{shock}$ is
called the shock front, and the curve in the range $-c_s \le \tilde x <
-v_\text{r-p}$ is called the rarefaction wave.

\begin{figure}
    \centering
     \begin{tikzpicture}[yscale=0.3,xscale=3,domain=-1:1]
        \def\emax{10}
        \def\emin{1}
        \def\cs{sqrt(1/3)}
        \def\vrp{0.1555959}
        \def\ep{3.139831558}
        \def\vs{0.75211538}

      \draw[thick,->] (-1.1,0) -- (1.1,0) node[right] {$\tilde{x}$};
      \draw[thick,->] (0,-0.2) node[below] {$0$} -- (0,11) node[below left] {$\tilde\epsilon(\tilde{x})$};

      \draw[thick,color=blue] ({-1},{\emax})-- ({-\cs},{\emax});
      \draw[thick,color=blue] ({-\vrp},{\ep})-- ({\vs},{\ep}) -- ({\vs},{\emin}) -- ({1},{\emin});
      \draw[thick,color=blue,domain={-\cs}:{-\vrp}] plot (\x,{\emax*((1-\cs)/(1+\cs)*(1-\x)/(1+\x))^((1+\cs*\cs)/(2*\cs))});

      \draw[thick,color=blue,dotted] ({-1.1},{\emax}) node[left,color=black] {$\epsilon_\mathrm{max}$} -- ({-1},{\emax});
      \draw[thick,color=blue,dotted] ({-1.1},{\ep}) node[left,color=black] {$\epsilon_\mathrm{plateau}$} -- ({-\vrp},{\ep});
      \draw[thick,color=blue,dotted] ({-1.1},{\emin}) node[left,color=black] {$\epsilon_\mathrm{min}$} -- ({\vs},{\emin});

      \draw[thick,color=blue,dotted] ({-\cs},{\emax}) -- ({-\cs},{0});
      \draw[thick,color=blue,dotted] ({-\vrp},{\ep})  -- ({-\vrp},{0});
      \draw[thick,color=blue,dotted] ({\vs},{\emin})  -- ({\vs},{0});

      \draw[thick] ({-\cs},{0})  -- ({-\cs},{-0.2})  node[below] {$-c_s$};
      \draw[thick] ({-\vrp},{0}) -- ({-\vrp},{-0.2}) node[below,xshift=-0.25cm] {$-v_\text{r-p}$};
      \draw[thick] ({\vs},{0})   -- ({\vs},{-0.2})   node[below] {$v_\mathrm{shock}$};
      \draw[thick] (-1,0) -- (-1,-0.2) node[below] {$-1$};
      \draw[thick] (1,0) -- (1,-0.2) node[below] {$1$};

    \end{tikzpicture}
    \caption{Analytical solution to the Riemann problem for $t>0$ with $\tilde x=x/t$.}
    \label{fig:riemannsol}
\end{figure}
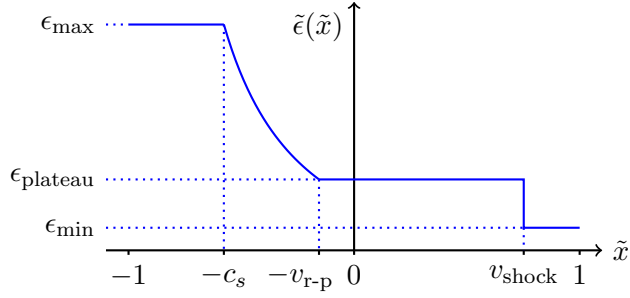

In heavy-ion collisions, the created matter is surrounded by the vacuum, where
$\epsilon(t,x)=0$, and the medium expands into it. Such a situation can be
mimicked in the Riemann problem by setting $\epsilon_\text{min}=0$, where the
solution is simplified to
\begin{align}
    \Tilde{\epsilon}(\Tilde{x}) &= \begin{cases}
       \epsilon_\text{max}, & \Tilde{x} < -c_s,\\
       \epsilon_\text{max}\left(\frac{(1-c_s)(1-x)}{(1+c_s)(1+x)}\right)^{\frac{1+c_s^2}{2 c_s}}, & -c_s \leq \Tilde{x} < 1,\\
       0, & 1\leq \Tilde{x}.
    \end{cases}
\end{align}

\subsubsection{Gubser flow}
\label{sec:Gubser}

Another exact solution is the Gubser flow~\cite{Gubser}, which is a solution
for the conformal EOS, $P(\epsilon)=\epsilon/3$, characterized by the boost
invariance and the \textit{azimuthal symmetry} (i.e., the rotational symmetry
in the $x$--$y$ plane). The solution is given in the Milne coordinates as
\begin{align}
    \epsilon(\tau,r) &= \frac{\epsilon_0 (2q)^{8/3}}{\tau^{4/3}\left(1+2q^2(\tau^2+r^2)+q^4(\tau^2-r^2)^2\right)^{4/3}}, \label{gubser}\\
    u^\tau(\tau,r) &= \cosh{k(\tau,r)},\\
    u^x(\tau,r) &= \frac{x}{r}\sinh{k(\tau,r)},\\
    u^y(\tau,r) &= \frac{y}{r}\sinh{k(\tau,r)}, \\
    u^\eta(\tau,r) &= 0,
\end{align}
with
\begin{align}
    k(\tau,r) &= \tanh^{-1}{\frac{2q^2\tau r}{1+q^2\tau^2+q^2r^2}},\\
    r &= \sqrt{x^2+y^2},
\end{align}
where $q$ and $\epsilon_0$ are free parameters.

\section{Discretization of hydrodynamic equations}
\label{sec:discretization}
To solve the hydrodynamic equations numerically, we must discretize both the
spatial and temporal coordinates. In this section, we first give a brief review
of the Runge-Kutta (RK) method with a focus on the differences between implicit
and explicit methods in Sec.~\ref{sec:RK}. The space discretization scheme and
the boundary condition applied in our study are then discussed in
Secs.~\ref{sec:space} and~\ref{sec:boundary}.

\subsection{Implicit and explicit Runge-Kutta methods}
\label{sec:RK}

The RK method is a common way to numerically solve a system of ordinary
differential equations (ODEs) of the general form,
\begin{equation}
    \partial_t \vec{y}(t) = \vec{h}(t,\vec{y}(t)), 
\end{equation}
where $\vec{y}(t)$ represents the full set of state variables.  The RK method
solves the variables $\vec{y}(t_\text{end})$ at the end time $t=t_\text{end}$
given an initial condition $\vec{y}(t_\text{init})$ at the initial time
$t=t_\text{init}$ by discretizing the time into steps of length $\Delta t$ and
iteratively updating $\vec{y}(t)$ for every time step.  For one time step from
$t$ to $t+\Delta t$, the state variable of the next time $\vec{y}(t+\Delta t)$
is constructed from $\vec{y}(t)$ as
\begin{align}
    \vec{y}(t+\Delta t) &= \vec{y}(t) + \Delta t\sum^S_{n=1} b_n \vec{k}_{(n)}, \label{nexttime}
    \\
    \vec{k}_{(n)} &= \vec{h}\Bigg(t+c_n\Delta t, \vec{y}(t)  + \Delta t\sum^S_{m=1} a_{nm} \vec{k}_{(m)}\Bigg), \label{ki}
\end{align}
where $S$ is the number of stages of the RK method, and the coefficients
$a_{nm}$, $b_n$, $c_n$ for $n,m=1,\cdots, S$ are determined so that the method
has desired properties. One specific set of the coefficients defines a
particular RK method.  The table that summarizes these coefficients is called a
Butcher table, whose examples are shown in Table~\ref{tab:butcher}.

In a RK method, the numerical error of $\vec{y}(t_\text{end})$ from the exact
solution $\vec{y}^*(t_\text{end})$ is suppressed in the limit $\Delta t\to0$ as
\begin{equation}
    \|\vec{y}(t_\text{end}) - \vec{y}^*(t_\text{end})\| \leq C(t_\text{end}) {\Delta t}^p,
    \label{accuracyOrder}
\end{equation}
where $C(t)$ is specific to the ODE and is independent of $\Delta t$. The
largest integer $p$ satisfying Eq.~\eqref{accuracyOrder} is called the order of
accuracy that is specific to each RK method.

The RK methods are classified into two categories: explicit and implicit
ones. A RK method is said to be explicit when $a_{nm}=0$ for $m\ge n$. In this
case, $\vec{k}_{(n)}$ in Eq.~\eqref{ki} depend only on $\vec{y}(t)$ and
$\vec{k}_{(m)}$ for $m<n$, so $\vec{k}_{(n)}$ are sequentially obtained from
$n=1$ to $S$ by simply substituting the known values. The other RK methods are
said to be implicit. For implicit methods, one must solve the set of equations
Eq.~\eqref{ki} for $\vec{k}_{(n)}(t)$, which is typically nonlinear and
practically carried out by a numerical solver. The implicit methods are thus
usually considered to be computationally more expensive.

Despite this disadvantage, the implicit methods have a useful property, i.e.,
the implicit methods are always linearly stable (the $A$-stability) as far as
the exact solution is not divergent~\cite{radau}. This means that a larger
$\Delta t$ can be adopted for stiff equations in implicit methods while keeping
numerical error finite. Another property of the implicit methods is that they
can reach a given order of accuracy in fewer stages than explicit ones. More
specifically, explicit methods require as many stages as the order up to $p=4$
and more stages than the order for $p>4$. On the other hand, the maximum
accuracy for a given $S$ is $p=2S$ in implicit
methods~\cite{NumericalAnalysis}.

\begin{table*}
    \centering
    \begin{tabular}{c|ccc}
         & Heun & Midpoint & Gauss-Legendre~1 (GL1)\\
        \hline \\[-0.5em]
        Type & Explicit & Explicit & Implicit \\
        \\
        Stage $S$ & $2$ & 2 & $1$ \\
         Order $p$ & 2 & 2 & 2\\
         \\
        \begin{tabular}{c|c}
            \multicolumn{2}{c}{} \\
            $c_n$ & $a_{nm}$\\
            \hline
                & $b_m$
        \end{tabular} &
        \begin{tabular}{c|cc}
          0 & 0 & 0 \\
          1 & 1 & 0 \\
         \hline
                  & 0.5 & 0.5
         \end{tabular} &
         \begin{tabular}{c|cc}
          0 & 0 & 0 \\
          0.5 & 0.5 & 0 \\
         \hline
            & 0 & 1
         \end{tabular} &
         \begin{tabular}{c|c}
            \multicolumn{2}{c}{} \\
             0.5 & 0.5 \\
             \hline
               & 1
         \end{tabular} \\
    \end{tabular}
    \caption{Properties and the Butcher tables of Heun's, the midpoint, and GL1 methods.}
    \label{tab:butcher}
\end{table*}

\subsection{Space discretization}
\label{sec:space}

In computational hydrodynamics, we discretize the spatial coordinates to
represent the state of the fluid using a finite number of degrees.  The spatial
derivatives are then approximated in terms of discretized fields so that the
partial differential equations become a set of ODEs, which can be solved by a
RK method.  This discretization has to be carefully chosen to suppress
numerical diffusion but at the same time to avoid spurious oscillations.

In this study, we employ for the space discretization the Kurganov-Tadmor (KT)
scheme with the MUSCL (monotonic upstream-centered scheme for conservation
laws) reconstruction~\cite{KT}. This method has been applied to relativistic
hydrodynamics for HIC in Refs.~\cite{Schenke:2010nt, Pang:2018zzo}. The KT
method has second-order accuracy for space discretization and is designed to
capture discontinuities accurately and suppress spurious oscillations. Another
important property of the KT scheme for our study is its independence of time
discretization. This property allows us to combine this scheme with arbitrary
time integration methods and compare them.

To discretize the space in the Cartesian coordinates, we assume a structured
grid where the space is discretized into cells by an equal mesh size $\Delta x$
along each coordinate.  In the KT scheme, one chooses the set of the conserved
densities of the cells, $\{[T^{t\nu}]_j\}$, as the state variables, where the
subscript $j$ of the square bracket specifies a cell. We spatially discretize
the continuity equation
\begin{align}
    \partial_t T^{t\nu} &= -\partial_xT^{x\nu} -  \partial_yT^{y\nu} - \partial_zT^{z\nu}, &
    (\nu=t,x,y,z),
    \label{hydroeq}
\end{align}
as
\begin{align}
    \partial_t [T^{t\nu}]_j
    &= [h_\text{KT}(T^{t\nu})]_j,
    \label{hydroKT}
\end{align}
with
\begin{equation}
    [h_\text{KT}(T^{t\nu})]_j = - [\text{KT}_x(T^{t\nu})]_j - [\text{KT}_y(T^{t\nu})]_j - [\text{KT}_z(T^{t\nu})]_j, 
\end{equation}
using the KT operators $\text{KT}_i$. In our case, the KT operator in the $x$
direction, $\text{KT}_x$, is written as
\begin{align}
    [\text{KT}_x(T^{t\nu})]_j =& \frac{H_{j+1/2}-H_{j-1/2}}{\Delta x}, \\ 
    H_{j+1/2} =& \frac{T^{x\nu}\big([T^{t\nu}]^+_{j+1/2}\big)+T^{x\nu}\big([T^{t\nu}]^-_{j+1/2}\big)}{2}
    \notag \\
    &+ \frac{a_{j+1/2}}{2}\left([T^{t\nu}]^+_{j+1/2}-[T^{t\nu}]^-_{j+1/2}\right),
    \label{H_j+1/2}
\end{align}
where $H_{j+1/2}$ is the flux at the boundary, $H_{j-1/2} = H_{(j-1)+1/2}$ with
$j\pm n$ representing the $\pm n$-th neighboring cell in the $x$ direction from
$j$, and $[T^{t\nu}]^\pm_{j+1/2}$ are values reconstructed on the cell boundary
that will be specified below.  The function $T^{x\nu}([T^{t\nu}]^\pm_{j+1/2})$
calculates the flux as a function of the reconstructed $[T^{t\nu}]^\pm_{j+1/2}$
using Eqs.~\eqref{eq:determine-ev} and~\eqref{u0i}.  In this study, we use the
Newton method to solve Eq.~\eqref{eq:determine-ev} for $v$.  The maximum local
speed of propagation is given by
\begin{equation}
    a_{j+1/2} = \max\left\{\rho\Bigl(J_{\alpha\beta}\bigl([T^{t\nu}]^+_{j+1/2}\bigr)\Bigr),\rho\Bigr(J_{\alpha\beta}\bigl([T^{t\nu}]^-_{j+1/2}\bigr)\Bigr)\right\},
\end{equation}
where $J_{\alpha\beta}(T^{t\nu}) = \partial T^{x\alpha}/\partial
T^{t\beta}|_{T^{t\nu}}$ is the $4\times4$ Jacobian matrix, and $\rho(M)$ is the
spectral radius of a matrix $M$~\footnote{An example of the analytic formula of
$\rho(J_{\alpha\beta}(T^{t\nu}))$ can be found in Ref.~\cite{Schenke:2010nt}.}.

In the original study of the KT scheme~\cite{KT}, the quantities on the cell
boundary $Q^\pm_{j+1/2}$ are calculated from the cell values $Q_j$ using the
MUSCL reconstruction as
\begin{align}
Q^\pm_{j+1/2} &= Q_{j+1/2\pm1/2} \mp\frac{\Delta x}{2}[\hat\partial_x Q]_{j+1/2\pm1/2}, \label{Tpm} \\
     [\hat\partial_x Q]_{j} &= \text{minmod}\Bigg(\text{minmod}\Bigg(\theta\frac{Q_j-Q_{j-1}}{\Delta x}, \frac{Q_{j+1}-Q_{j-1}}{2\Delta x}\Bigg),\theta\frac{Q_{j+1}-Q_j}{\Delta x}\Bigg),
     \label{dTpm} \\
    \text{minmod}(a,b) &= \frac{\text{sgn}(a)+\text{sgn}(b)}{2}\min(|a|, |b|). \label{Tpmminmod}
\end{align}
The flux limiter in Eq.~\eqref{dTpm} guarantees the non-oscillatory property of
$Q$ with the parameter $\theta$ restricted to $1\leq\theta\leq 2$. Choosing
$\theta < 1$ would produce excessive numerical diffusion, whereas choosing
$\theta > 2$ would produce unphysical oscillation.  We choose $\theta=1.1$
following Ref.~\cite{Schenke:2010nt}, which gives non-oscillatory results with
a limited numerical diffusion. We verified that changing the value of $\theta$
hardly changes the tendency of our results for accuracy and efficiency in
Sec.~\ref{sec:results} unless $\theta$ is too close to two.

In relativistic hydrodynamics, the reconstruction could be simply applied by
choosing $Q=T^{t\nu}$ to directly obtain the boundary values
$[T^{t\nu}]^\pm_{j+1/2}$ in Eq.~\eqref{H_j+1/2} from $[T^{t\nu}]_j$.  However,
to enhance the numerical stability of our implementation, we instead choose $Q
= (m, T^{ti})$ for the spatial reconstruction, where $m$ is the mass density of
the fluid cell defined by
\begin{align}
    m^2 = g_{\mu\nu}T^{t\mu}T^{t\nu} = g_{tt} (T^{tt})^2 - K^2. \label{m^2}
\end{align}
More explicitly, we obtain $[T^{t\nu}]^\pm_{j+1/2}$ as follows.

First, a set of conserved densities $[T^{t\nu}]_j$ obtained in each
intermediate numerical step can fail in reproducing the physical requirement
$m^2\ge0$.  To avoid this situation, whenever a set of $[T^{t\nu}]_j$ is
calculated, we replace $T^{tt}$ with
\begin{equation}
    \max\left(T^{tt},(1+s)K\right),
\end{equation}
where a small regulator $s$ is chosen to be $10^{-15}$ for the present study.

Second, the MUSCL reconstruction would guarantee the non-oscillatory property
of the reconstructed variables $Q$. We confirmed this property when
Eqs.~\eqref{Tpm} and~\eqref{dTpm} are applied to $Q=T^{t\nu}$. However, in this
procedure, the non-oscillatory property of $m^2$ is not guaranteed, and thus
the requirement $m^2>0$ can be broken. As a practical matter, we are faced with
unphysical finite oscillations in $\epsilon$ and $u^\mu$ obtained from
$T^{t\nu}$ of this reconstruction.  To suppress these oscillations, we instead
apply the reconstruction by the choice $Q=(m, T^{ti})$ as in the following
procedure:
\begin{enumerate}
    \item
    Starting from $[T^{t\nu}]_j$, construct the local mass density $[m]_j$ using Eq.~\eqref{m^2}.
    \item
    Reconstruct $[m]^\pm_{j+1/2}$ and $[T^{ti}]^\pm_{j+1/2}$ from $[m]_j$ and $[T^{ti}]_j$, respectively, using Eq.~\eqref{Tpm}.
    \item
    Determine $[T^{tt}]_{j+1/2}^\pm$ from $[m]_{j+1/2}^\pm$ and $[T^{ti}]_{j+1/2}^\pm$ by inverting Eq.~\eqref{m^2}.
\end{enumerate}
We confirmed that this prescription significantly reduces the oscillating
behavior of $\epsilon$ and $u^\mu$ in our numerical tests.

The KT operators in the other directions, $\text{KT}_y$ and $\text{KT}_z$, are
defined similarly.  The right-hand side of Eq.~\eqref{hydroKT} combined with
the MUSCL reconstruction~\eqref{Tpm}--\eqref{Tpmminmod} references two cells
away from the current cell in every direction.  This property plays an
important role in optimizing the implicit solver discussed below.

In the Milne coordinates, we solve hydrodynamics equations in the form of
Eqs.~\eqref{eq:milnet} and~\eqref{eq:milnel} with $\tilde{T}^{t\nu}$ as the
state variables. In this case, we use the same prescription as above with the
following modifications: To obtain the energy density and velocity,
Eqs.~\eqref{eq:determine-ev} and~\eqref{u0i} are applied after transforming
$T^{\tau \nu}$ to the local orthonormal frame, and the result is transformed
back to the Milne basis.  In the spatial reconstruction, we use $\tilde{m} =
\tau m$ in place of the mass density $m$.  The geometric source term
$P(\epsilon)$ in Eq.~\eqref{eq:milnet} is added to the right-hand side of
Eq.~\eqref{hydroKT}.

\subsection{Boundary conditions and vacuum}
\label{sec:boundary}

Since the numerical evolution of hydrodynamics is performed on a finite grid,
one has to introduce boundary conditions. In our space scheme the evaluation of
$[\text{KT}_i]_j$ references two neighboring cells. Therefore, the variables on
these cells must be specified when they are outside the numerical grid. In this
study, we employ the same boundary condition as that used in
Ref.~\cite{Karpenko:2013wva}.  In this boundary condition, when cells $j\pm1$
are outside the numerical grid, we use the variables on cell $j$ for those of
cell $j\pm1$. If cells $j\pm2$ are outside the numerical grid, the variables on
cells $j\pm1$ are substituted there.  Since this boundary condition roughly
approximates an outflow without reflection, it is sufficient for an application
to heavy-ion collisions when the grid is large enough that the boundary is
essentially kept in the vacuum.

In the numerical analysis in Sec.~\ref{sec:results}, we deal with the vacuum
region with $\epsilon(\bm{x}_j) = 0$, where $\bm{x}_j$ represents the spatial
coordinates of cell $j$. However, the treatment of $\epsilon(\bm{x}_j)=0$ is
problematic since it leads to an undefined velocity through Eq.~\eqref{u0i}. To
avoid this situation, we force the energy-density $\epsilon(\bm{x}_j)$ to be
always at least $\epsilon_\text{vac} = 10^{-100}~\text{fm}^{-4}$ in
Eq.~\eqref{eq:determine-ev}.

\section{Locally optimized fixed-point implicit solver}
\label{sec:time}

The time evolution of the differential equations~\eqref{hydroKT}, which are
obtained by the space discretization of Eq.~\eqref{hydroeq}, can be solved by
RK methods introduced in Sec.~\ref{sec:RK}, where $\vec{y}(t)$ and
$\vec{h}(\vec{y})$ in Eq.~\eqref{ki} correspond to the state variables
$\{[T^{t\nu}]_j\}$ and the right-hand side of Eq.~\eqref{hydroKT},
respectively. For implicit RK methods, we need to numerically solve implicit
equations for substantial problems. However, implicit methods usually cannot be
efficient without modifications compared to explicit ones.

To overcome this excessive computational cost of implicit methods, we propose
an improved solver for implicit RK methods in this study. After defining our
method in Sec.~\ref{sec:implsolving}, we discuss the convergence condition of
our method and its connection to the stiffness and stability in
Sec.~\ref{sec:stiffness}. In Sec.~\ref{sec:AB2}, we finally discuss the
accuracy, efficiency, and stability for a specific case of the implicit RK
method called the one-stage Gauss-Legendre (GL1) method by relating it to the
two-step Adams-Bashforth (AB2) method.

\subsection{Fixed-point solver with local optimization}
\label{sec:implsolving}

\begin{figure}
    \centering
    \begin{tikzpicture}[declare function={
            f(\x) = 1+2/(1+exp(1.5*(\x-2)));
        }]
        \def\xa{0.5}
        \def\xb{f(\xa)}
        \def\xc{f(\xb)}
        \def\xd{f(\xc)}
        \def\xe{f(\xd)}
        \def\xf{f(\xe)}
        \def\xg{f(\xf)}
        \def\xh{f(\xg)}
        \def\xinf{f(f(f(f(f(f(f(\xg)))))))}

      \draw[thick,->] (-0.2,0) -- (4.2,0) node[right] {$K$};
      \draw[thick,->] (0,-0.2) -- (0,4.2) node[above] {$y$};
      \draw[thick,-,dashed] (-0.1,-0.1) -- (4.1,4.1) node[above,xshift=0.2cm] {$y = K$};
      \draw[thick,color=red] ({\xa},0) node[below] {$K^{(0)}$}
        -- ({\xa},{\xb}) -- ({\xb},{\xb}) 
        -- ({\xb},{\xc}) -- ({\xc},{\xc})
        -- ({\xc},{\xd}) -- ({\xd},{\xd})
        -- ({\xd},{\xe}) -- ({\xe},{\xe})
        -- ({\xe},{\xf}) -- ({\xf},{\xf})
        -- ({\xf},{\xg}) -- ({\xg},{\xg})
        -- ({\xg},{\xh})
        ;
      \draw[thick,color=red,dotted] ({\xb},{\xc}) -- ({\xb},0) node[below,xshift=0.2cm] {$K^{(1)}$};
      \draw[thick,color=red,dotted] ({\xc},{\xc}) -- ({\xc},0) node[below] {$K^{(2)}$};
      \draw[thick,color=red,dotted] ({\xd},{\xe}) -- ({\xd},0) node[below] {$K^{(3)}$};
      \draw[thick,color=black,dotted] ({\xinf},{\xinf}) -- ({\xinf},0) node[below] {$K^{*\phantom{()}}$};

      \draw[thick,color=blue,domain=-0.2:4.2] plot (\x,{f(\x)}) node[above,xshift=-0.2cm,yshift=0.2cm] {$y=F(K)$};
    \end{tikzpicture}
    \caption{Schematic picture of the iterative process $K^{(0)}$, $K^{(1)}, \cdots$ in the fixed-point method toward the solution $K^*$ for the single-variable case.}
    \label{fig:fix}
\end{figure}
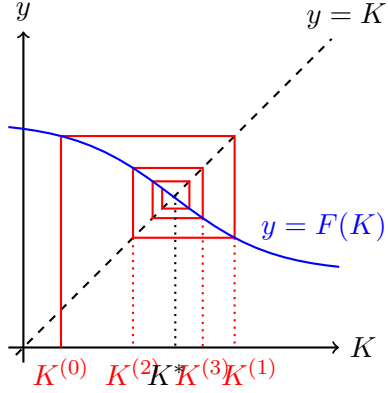

In implicit RK methods, $\vec{k}_{(n)}$ in Eq.~\eqref{ki} are obtained by
solving the system of implicit equations. In this study we adopt the
fixed-point method as a numerical solver for this procedure. To illustrate this
method, we first rewrite Eq.~\eqref{ki} in a compact form
\begin{align}
    \vec{K} = \vec{F}(\vec{K}),
    \label{K=FK}
\end{align}
with $\vec{K}=(\vec{k}_{(1)}, \cdots, \vec{k}_{(S)})$ and

\begin{align}
    \vec{F}(\vec{K}) &= (\vec{f}_{(1)}(\vec{K}), \cdots, \vec{f}_{(S)}(\vec{K}) ),
    \label{F_K}\\
    \vec{f}_{(n)}(\vec{K}) &= \vec{h}\Bigg(t + c_n\Delta t, \vec{y}(t) + \Delta t\sum_m a_{nm} \vec{k}_{(m)}\Bigg).
    \label{fofk}
\end{align}
The fixed-point method solves Eq.~\eqref{K=FK} by starting from an initial
guess $\vec{K}^{(0)}$ and iteratively updating the guess as
$\vec{K}^{(l+1)}=\vec{F}(\vec{K}^{(l)})$ for $l=0,1,2,\cdots$ until the
convergence is reached. A schematic illustration of this procedure for a
single-variable case is shown in Fig.~\ref{fig:fix}.

The convergence in the fixed-point method strongly depends on the choice of the
initial guess $\vec{K}^{(0)}$. The initial guess should not be too far from the
true solution. In this study, for the initial guess, we use the solution of
$\vec{K}$ in the previous RK step. This choice comes from an expectation that
$\vec{K}$ only changes at the order of $\mathcal{O}(\Delta t$), and hence
$\vec{K}$ at the previous RK time step is a good guess for the next one. As we
will see below, this choice of $\vec{K}^{(0)}$ plays a crucial role in
realizing a good efficiency in our method.  For the first time step of RK at
$t=t_\text{init}$, we use $\vec{K}^{(0)}=(0,\cdots,0)$ for the initial guess.

To improve the efficiency, we utilize the locality of discretized partial differential equations , i.e., a
finite number of cells are referenced in an evaluation of the time derivative
for one cell.  In our case,
the right-hand side of Eq.~\eqref{hydroKT} depends only on
cells up to two neighbors in each direction.  By taking this advantage, in our
implementation, we check the convergence cell by cell using the criterion
\begin{equation}
    \big\| [\vec{K}^{(l+1)}]_{j} - [\vec{K}^{(l)}]_j \big\| = \big\| [\vec{F}(\vec{K}^{(l)})]_j - [\vec{K}^{(l)}]_j  \big\| < e \frac{\langle T^{tt}\rangle}{\Delta t} \bigg(\frac{\Delta t}{\Delta x}\bigg)^p, \label{KFK<e}
\end{equation}
with a pre-determined parameter $e$, where $[\vec{K}]_j$ represents the set of
variables in $\vec{K}$ at cell $j$, and $\langle T^{tt}\rangle$ is the spatial
average of $T^{tt}$ in the entire numerical grid.  Note that
$[\vec{K}]_j$ includes $4S$ variables corresponding to $T^{t\nu}$ for all
stages.

Since $\langle T^{tt}\rangle/\Delta t$ has the same unit as
$[\vec{K}^{(l)}]_j$, $e$ is dimensionless. This parameter should be properly
tuned for each calculation setup since smaller $e$ leads to better accuracy but degrades the efficiency,
and vice versa.
In our numerical analysis in Sec.~\ref{sec:results} we choose
$e=10^{-3}$ and $e=2\times 10^{-4}$ for one and two space dimensions,
respectively.

In Eq.~\eqref{KFK<e}, we require that the threshold scales as $\Delta
t^{p-1}$. This $\Delta t$ dependence is necessary to preserve the accuracy order $p$ of the numerical results. Since the error of $[\vec{K}^{(l+1)}]_j$ is
parametrically of order $\Delta t$ higher than that of $[\vec{K}^{(l)}]_j$, the
error of $[\vec{K}^{(l+1)}]_j$ is of order $\Delta t^p$ when the
condition~\eqref{KFK<e} is met. This order of $[\vec{K}^{(l+1)}]_j$ is in
accordance with the accuracy of RK in Eq.~\eqref{accuracyOrder}~\footnote{The
numerical error of one RK step is of order $\Delta t^{p+1}$. Also, notice
$\Delta t$ on the right-hand side of Eq.~\eqref{nexttime}.}. On the right-hand side of Eq.~\eqref{KFK<e}, $\Delta x$ is used to
make $p$ dependence dimensionless.  In
Sec.~\ref{sec:results} we show that the expected scaling~\eqref{accuracyOrder}
is nicely obtained using the criterion~\eqref{KFK<e}.

More specifically, we update all the cells for the first iteration. For the
succeeding iterations, we update cell $j$ only when any of the following
conditions are met:
\begin{enumerate}
    \item Eq.~\eqref{KFK<e} was not satisfied at cell $j$ in the last update.
    \item Any of the cells adjacent to cell $j$ was updated in the previous
    iteration and did not satisfy Eq.~\eqref{KFK<e}.
\end{enumerate}
We update all the cells that need updating at once. We proceed with the
iterations until Eq.~\eqref{KFK<e} is reached for all the cells. We note that
even once the condition~\eqref{KFK<e} is satisfied at a cell, the condition can
be violated later due to updates of neighboring cells.

As we will show in the next section, this optimization drastically reduces the
computational cost. In fact, only a single update is enough for most cells. We
refer to this prescription as the \textit{local optimization} in the following.

In the above procedure, we check only the most adjacent cells. Although the
right-hand side of Eq.~\eqref{hydroKT} also depends on the second neighboring
cells, we have checked that their inclusion hardly changes the quality of
numerical results despite an extra computational cost~\footnote{The success of this prescription may
rely on the fact that Eq.~\eqref{hydroKT} only depends on at most second
neighboring cells. When one employs a space scheme that depends on more distant
cells, this procedure may need to be modified to keep the quality of numerical
results.}. We also notice that an
update of a cell does not necessarily lead to the update of its neighboring
cells in the next iteration.

Strictly speaking, the local optimization violates the energy--momentum
conservation.  This is because the flux at a boundary between two cells can be
inconsistent when one cell is updated but the other is not.  The numerical
error from this violation, however, is controlled within the accuracy as we
will show in Sec.~\ref{sec:results}.

\subsection{Convergence condition and stiffness}
\label{sec:stiffness}

As is common in iterative methods, the fixed-point method does not necessarily
converge to the true solution.  A necessary condition for the convergence of
the fixed-point method is written in terms of the spectral radius $R$ of the
Jacobian matrix of $\vec{F}$ at the solution $\vec{K}=\vec{K}^*$:
\begin{equation}
   R = \rho\Bigg(\frac{\partial \vec{F}}{\partial \vec{K}}(\vec{K}^*)\Bigg) < 1, \label{rhof}
\end{equation}
Moreover, the convergence is faster for smaller $R$. From Eqs.~\eqref{F_K}
and~\eqref{fofk} one finds that the Jacobian matrix $\partial \vec{F}/\partial
\vec{K}$, and hence $R$, is proportional to $\Delta t$.  Therefore, the
convergence is guaranteed for a sufficiently small $\Delta t$ and is faster for
a smaller $\Delta t$.  However, the convergence is not guaranteed in
general. In fact, we will see later that our method can be aborted for $\Delta
t/\Delta x\gtrsim0.4$ due to a failure of the fixed-point solver. The
convergence in these cases can be established with additional prescriptions as
will be discussed in the forthcoming publications~\cite{future}.  In the
present study, we simply neglect these cases since our method works stably in
the range of $\Delta t$ which is practically important.

The convergence condition~\eqref{rhof} is related to the stiffness of the
system of equations. In fact, Eq.~\eqref{rhof} is rewritten as
\begin{align}
   R
   &= \rho\Biggg(\Delta t\Biggg[a_{nm} \frac{\partial\vec{h}}{\partial\vec{y}}\Biggr|_{(t+c_n\Delta t,\vec{y}+\Delta t \sum_{l=1}^S a_{nl} \vec{k}^*_{(l)})}\Biggg]_{nm}\Biggg) \nonumber \\
   &= \Delta t \rho(a_{nm})\rho\Biggl(\frac{\partial\vec{h}(t,\vec{y})}{\partial\vec{y}}\Biggr) + \mathcal{O}\bigl((\Delta t)^2\bigr),
   \label{eq:Rstiffness}
\end{align}
where $\rho(a_{nm})$ is the spectral radius of the Butcher table of a given RK
method. Here, $\rho(\partial \vec{h}/\partial \vec{y})$ characterizes the
stiffness of the system since the largest absolute eigenvalue $\max_\lambda
|\lambda| = \rho(\partial \vec{h}/\partial \vec{y})$ (i.e., the inverse of the
shortest timescale) gives the most severe impact on the
stability. Equation~\eqref{eq:Rstiffness} shows that $R$ is proportional to the
stiffness $\rho(\partial \vec{h}/\partial \vec{y})$.

This means that the convergence becomes slow in stiff systems. Since we locally
truncate the iteration number of the fixed-point method using
Eq.~\eqref{KFK<e}, the iteration number becomes automatically larger in the
stiff spatial region where short-timescale phenomena are happening. In short,
our local optimization automatically detects the stiffness of local regions.

\subsection{Accuracy, efficiency, and stability}
\label{sec:AB2}

In later sections, we particularly consider the one-stage Gauss-Legendre (GL1)
method~\cite{Kuntzmann:1961-GL, Butcher:1964-GL, NumericalAnalysis} as an
implicit method of order $p=2$, whose Butcher table is given in the right
column of Table~\ref{tab:butcher}. In this subsection, we focus on the locally
optimized GL1 and discuss its accuracy, efficiency, and stability.

We first show that the method applied to GL1 keeps the accuracy order $p=2$
even when the iteration number is fixed to one.  The single-iteration GL1 is
explicitly written down as
\begin{align}
    \vec y(t+\Delta t) &= \vec y(t) + \Delta t \vec k^{(1)},
    \label{eq:GLy}
    \\
    \vec k^{(1)} &= \vec h\Bigg( t + \frac{\Delta t}2 , \vec y(t)+ \frac{\Delta t}2 \vec k^{(0)} \Bigg),
    \label{eq:GLk}
\end{align}
where the initial guess $\vec k^{(0)}$ of the iteration is chosen to be the
solution in the previous time step.  By inverting Eq.~\eqref{eq:GLy} of the
previous time step, the initial guess can be expressed as
\begin{align}
  \vec k^{(0)} = \frac{\vec y(t)-\vec y(t-\Delta t)}{\Delta t}.
  \label{eq:GLinit}
\end{align}
Plugging Eqs.~\eqref{eq:GLk} and~\eqref{eq:GLinit} into Eq.~\eqref{eq:GLy},
we obtain
\begin{align}
  \vec{y}(t+\Delta t)
    &= \vec{y}(t) + \Delta t \vec{h}\Biggl(t + \frac{\Delta t}2, \frac{3\vec{y}(t) - \vec{y}(t-\Delta t)}2\Biggr).
    \label{eq:GL1.single-step.combined}
\end{align}
The Taylor expansion of this expression matches the analytic expansion up to
the second order in $\Delta t$:
\begin{equation}
  \vec{y}(t+\Delta t)
    = \vec{y}(t) + \Delta t \vec{h}\bigl(t, \vec{y}(t)\bigr)
    + \frac{(\Delta t)^2}2 [\partial_t + \vec{h}\bigl(t, \vec{y}(t)\bigr) \cdot \partial_{\vec{y}}] \vec{h}\bigl(t, \vec{y}(t)\bigr) + \mathcal{O}\bigl((\Delta t)^3\bigr).
\end{equation}
Thus, the single-iteration fixed-point method for GL1 already has the desired
accuracy of order $p=2$.

Next, we see that the single-iteration GL1 can be related to AB2.  By regarding
\begin{align}
    \vec{z}\ \bigg(t+\frac{\Delta t}2\bigg)
    =&\ \frac{3\vec{y}(t) - \vec{y}(t-\Delta t)}2
    \notag \\
    =&\ \vec{y}(t - \Delta t) + \frac32 \Delta t \vec{h}\Biggl(t-\frac{\Delta t}2,\vec{z}\biggl(t - \frac{\Delta t}2\biggr)\Biggr).
\end{align}
as state variables, Eq.~\eqref{eq:GL1.single-step.combined} can be shown to be
equivalent to the AB2 method:
\begin{equation}
  \vec{z}(t+\Delta t) = \vec{z}(t) + \frac32\Delta t \vec{h}\bigl(t,\vec{z}(t)\bigr) - \frac12\Delta t \vec{h}\bigl(t-\Delta t,\vec{z}(t-\Delta t)\bigr),
  \label{eq:GL1.single-step.AB2}
\end{equation}
Here, the initial guess $\vec{k}^{(0)}$ plays an important role in the
equivalence of the single-iteration GL1~\eqref{eq:GL1.single-step.combined} to
AB2~\eqref{eq:GL1.single-step.AB2}. In the single-iteration GL1, the initial
guess carries the information of the previous step $\vec y(t-\Delta t)$, which
is similar to the linear multistep methods such as AB2 using the states of
previous steps.

Both the single-iteration GL1 and AB2 are second-order methods with one-stage
evaluation of $\vec{h}$, so they are more efficient than the second-order
explicit RK with two stages. However, the problem with those one-stage methods
is the stability in the stiff equations. Those methods effectively use an
extrapolation using the previous step $\vec{y}(t - \Delta t)$ as observed in
Eqs.~\eqref{eq:GL1.single-step.combined} and~\eqref{eq:GL1.single-step.AB2} and
are known to be less stable than the explicit RKs. This is a common problem
with the linear multistep method such as the Adams-Bashforth method.  In our
method, this problem is automatically remedied by the iterations. As already
discussed, the number of fixed-point iterations increases in the stiff region
so that the method approaches the original GL1, which is $A$-stable. Thus, we
can view our method as automatic switching between the two second-order
methods, the less stable but efficient AB2 and the stable GL1, based on the
local stiffness. In this way, our method benefits from both the efficiency of
AB2 and the stability of GL1.

As we will observe in the numerical tests in the later sections, the stiff part
appears near the vacuum or the region with large gradients. Such regions
typically appear only in limited spacetime volumes in practical applications,
so we can expect the computational cost of the locally optimized implicit
method would be close to the cost of AB2\@. Since AB2 with a one-stage
evaluation of $\vec{h}\bigl(t,\vec{y}(t)\bigr)$ has half the cost of the
second-order explicit RK methods with two stages, we may expect a better
performance of the locally optimized implicit method as far as the total size
of the stiff regions is not too large.

The application of the idea of the locally optimized implicit method is not
limited to relativistic hydrodynamics. For example, the idea can be applied to
the general (discretized) field equations with local interactions. In
principle, this idea can be applied to any ODEs with the locality, where a
degree of freedom directly interacts with a fixed number of other degrees
regardless of the size of the entire system. Thus, we expect broader
applications of this idea in various fields.

Finally, we comment on the future extension of our analysis to the higher-order
implicit RK methods. In this subsection, we considered the specific case of GL1
utilizing its particular form. It should be in principle possible to design the
proper choice of the initial guess $\vec{K}^{(0)}$ for the general implicit RK
methods by taking the linear combinations of the solutions $\vec{K}$ in the
previous time steps so that the accuracy order of the implicit method is
preserved even with a single iteration. These analyses are left for future
study.

\section{Numerical setup}
\label{sec:setup}

\subsection{Runge-Kutta methods}

We have tested various implicit and explicit RK methods with different stages
and orders. They include the Gauss-Legendre family for the implicit methods,
and Heun's and the Euler methods for the explicit.  For the demonstration
purpose of our study, we choose Heun's method as the representative of explicit
methods, as it is commonly used in hydrodynamic codes for
HIC~\cite{Schenke:2010nt, Pang:2018zzo}. This method has two stages ($S=2$) and
second order of accuracy ($p=2$), and its Butcher table is given in the left
column of Table~\ref{tab:butcher}. For the implicit method, we choose the
one-stage Gauss-Legendre method~\cite{Kuntzmann:1961-GL, Butcher:1964-GL,
NumericalAnalysis}, which has the same order of accuracy as Heun's method. In
Sec.~\ref{sec:results}, we compare the numerical results of these methods. In
the following, we refer to Heun's and one-stage Gauss-Legendre methods as the
explicit and implicit methods, respectively.

Another conventional explicit method at the same order of accuracy as Heun's
method is the midpoint method, whose Butcher table is given in the middle
column of Table~\ref{tab:butcher}. In Sec.~\ref{sec:midpoint}, we show that the
performance of this method is almost the same as Heun's method in the Riemann
problem.

\subsection{Dimensions}

In this study, we compare the implicit and explicit RK methods in (1+1)- and
(2+1)-dimensional numerical simulations. For the (1+1)-dimensional case, we
deal with the Riemann problem discussed in Sec.~\ref{sec:Riemann} with finite
and zero minimum energy density $\epsilon_\text{min}$. We also perform
(2+1)-dimensional simulations with the boost invariance for the Gubser flow and
event-by-event initial conditions for heavy-ion collisions generated by
\TRENTo~\cite{Trento}.

In all cases, we consider a system of length 40~fm in each direction. To
investigate the dependence on the mesh size $\Delta x$, we perform numerical
simulations with $\Delta x = 0.2$ and $0.4~\text{fm}$ where the spatial extent
is divided into $200$ and $100$ cells, respectively.  The mesh size $\Delta x =
0.2~\mathrm{fm}$ is a typical value used in applications to 
collisions HIC~\cite{Schenke:2010nt,Hirano:2000xa,Karpenko:2013wva}. The value
$\Delta x = 0.4~\text{fm}$ is employed for comparison.  The time step $\Delta
t$ is fixed in a single simulation, and we perform simulations with different
$\Delta t$ down to $\Delta t = 0.1\times 2^{-5} \Delta x$, which is determined
so that our results in Sec.~\ref{sec:results} contain enough points to
investigate the accuracy order $p$ through Eq.~\eqref{accuracyOrder}.  We also
checked that our results behave consistently even at $\Delta t = 0.1\times
2^{-6} \Delta x$ and $0.1\times 2^{-7} \Delta x$ in several cases.

\subsection{Error estimate}
\label{sec:error-estimate}

The results obtained by computational hydrodynamics have discretization
errors. Numerical simulations are also affected by floating-point errors in
general. In the present study, we evaluate these errors using two quantitative
measures. First, we compare the energy density obtained by the numerical
analysis, $\epsilon_\text{num}(\bm{x}_j)$, with that of the exact solution
$\epsilon_\text{exact}(\bm{x}_j)$ in the continuum case using the measure
\begin{align}
    \Delta_\text{exact}(\bm{x}_j) = D\big(\epsilon_\text{num}(\bm{x}_j),\epsilon_\text{exact}(\bm{x}_j) \big) ,
    \label{Deltae_exact}
\end{align}
with
\begin{equation}
    D(\epsilon_1,\epsilon_2) = \frac{ \epsilon_1-\epsilon_2 }{\max(\epsilon_1,\epsilon_2)}.
    \label{D_e1_e2}
\end{equation}
Here, we take the maximum of $\epsilon_1$ and $\epsilon_2$ for the denominator
of Eq.~\eqref{D_e1_e2}, instead of either one of them, to avoid a situation
where Eq.~\eqref{Deltae_exact} becomes extremely large when one of them is
vanishingly small.
As for the second measure, we choose a numerical result obtained at small
$\Delta t$, denoted by $\epsilon_\text{ref}(\bm{x}_j)$, as a reference.  We
then compare $\epsilon_\text{num}(\bm{x}_j)$ with the reference as
\begin{align}
    \Delta_\text{ref}(\bm{x}_j) = D \big(\epsilon_\text{num}(\bm{x}_j),\epsilon_\text{ref}(\bm{x}_j)\big) .
    \label{Deltae}
\end{align}
For the reference $\epsilon_\text{ref}(\bm{x}_j)$, we use the result by the
explicit method with the smallest $\Delta t=0.1\times 2^{-5}\Delta x$.

To compare numerical errors of different time discretization methods,
$\Delta_\text{ref}(\bm{x}_j)$ is more convenient than
$\Delta_\text{exact}(\bm{x}_j)$.  This is because a finite space-discretization
error remains in $\Delta_\text{exact}(\bm{x}_j)$ even in the $\Delta t\to0$
limit, while one can discuss genuinely the time-discretization error using
$\Delta_\text{ref}(\bm{x}_j)$.

In Sec.~\ref{sec:results}, we inspect the magnitude of numerical error using
the maximum and spatial average of $|\Delta_\text{ref}(\bm{x}_j)|$. However, we
found that $|\Delta_\text{ref}(\bm{x}_j)|$ tends to be close to unity near the
vacuum region where $\epsilon_\text{exact}(\bm{x}_j)$ is vanishingly small. In
this region $\epsilon_\text{num}(\bm{x}_j)$ and $\epsilon_\text{ref}(\bm{x}_j)$
have non-vanishing values due to the numerical diffusion, i.e., the unphysical
propagation from the matter region, which is inevitable in computational
hydrodynamics. Nevertheless, $\epsilon_\text{num}(\bm{x}_j)$ is still
negligibly small and the large $|\Delta_\text{ref}(\bm{x}_j)|$ should not be
taken into account as an essential error.

To avoid this apparent error in the vacuum region, in our error analysis we
exclude the cells where $\epsilon_\text{num}(\bm{x}_j)$ is smaller than a
threshold value $\epsilon_\text{thr}$, where $\epsilon_\text{thr}$ is
determined so that the sum of the energy in the excluded cells is less than
$10^{-6}$ times the total energy.  We also exclude cells within $1\,\text{fm}$
from the boundaries to remove possible boundary effects. We then evaluate the
maximum and the average of $|\Delta_\text{ref}(\bm{x}_j)|$ in the remaining
cells.

\subsection{Estimate of computational cost}

To compare the performance of different numerical methods, we also need a
quantitative measure of the computational cost. However, the actual
computational cost depends on the implementation, and its appropriate
comparison is difficult in general. In this study, we focus on the fact that
the most computationally demanding part of our analysis is the evaluation of
the local KT operator $[h_\text{KT}]_j$ in Eq.~\eqref{hydroKT}. This means that
the number of evaluations $N_\text{KT}$ serves as an approximate measure of the
computational cost. We thus use the number of evaluations per cell,
\begin{equation}
    n_\text{KT} = \frac{N_\text{KT}}{N_\text{cell}},
    \label{cost}
\end{equation}
for the measure of the computational cost, where $N_\text{cell}$ stands for the
number of total cells. For our explicit method with $S=2$, $[h_\text{KT}]_j$ is
evaluated twice per cell in a single time step.  In the implicit method, the
cost of a single time step depends on the number of iterations of the
fixed-point solver, where one local iteration needs one evaluation of
$[h_\text{KT}]_j$ with $S=1$.  Moreover, due to the local optimization,
$n_\text{KT}$ is not necessarily an integer. We have checked that $n_\text{KT}$
indeed reflects the real computational times well with appropriate
optimizations.

\section{Numerical results}
\label{sec:results}

\subsection{1+1d Riemann problem with shock propagation}
\label{sec:result:Riemann}

Now, let us discuss the numerical results. We begin in this subsection to deal
with the Riemann problem, whose exact solution is presented in
Sec.~\ref{sec:Riemann}. This is a (1+1)-dimensional problem in flat spacetime
with initial conditions given by Eqs.~\eqref{Riemann:init}
and~\eqref{Riemann:init2}. We numerically solve the hydrodynamic evolution from
the initial time $t_\text{init}=0$ to $t_\text{end}=15~\text{fm}$. We set
$c_s^2=1/3$ and $\epsilon_\text{max} = 10~\text{fm}^{-4}$ and perform our
analysis for the two values of $\epsilon_\text{min}=1$ and
$0~\text{fm}^{-4}$. The latter is important for the application to heavy-ion
collisions where the matter expands into the vacuum, which typically causes
numerical problems.

\begin{figure}
    \centering
    \includegraphics[width=0.6\linewidth]{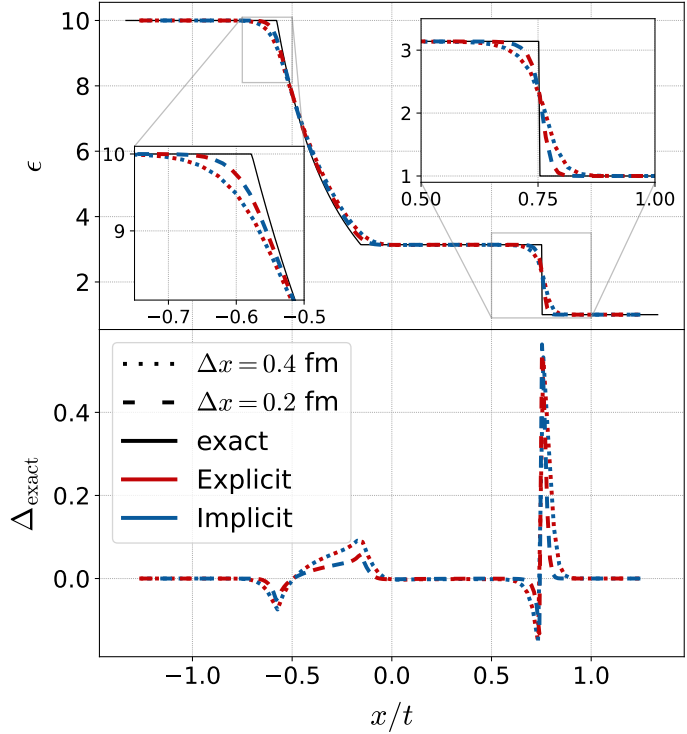}
    \caption{Energy density $\epsilon(\bm{x}_j)$ (upper panel) and its error compared with the exact solution $\Delta_\text{exact}(\bm{x}_j)$ (lower panel) for the Riemann problem at $t_\text{end}=15~\text{fm}$ with $\epsilon_\text{min}=1~\text{fm}^{-4}$ and $\Delta t = 0.1\times 2^{-5}\Delta x$.}
    \label{fig:riemannE}
\end{figure}

\begin{figure}
    \centering
    \includegraphics[width=0.6\linewidth]{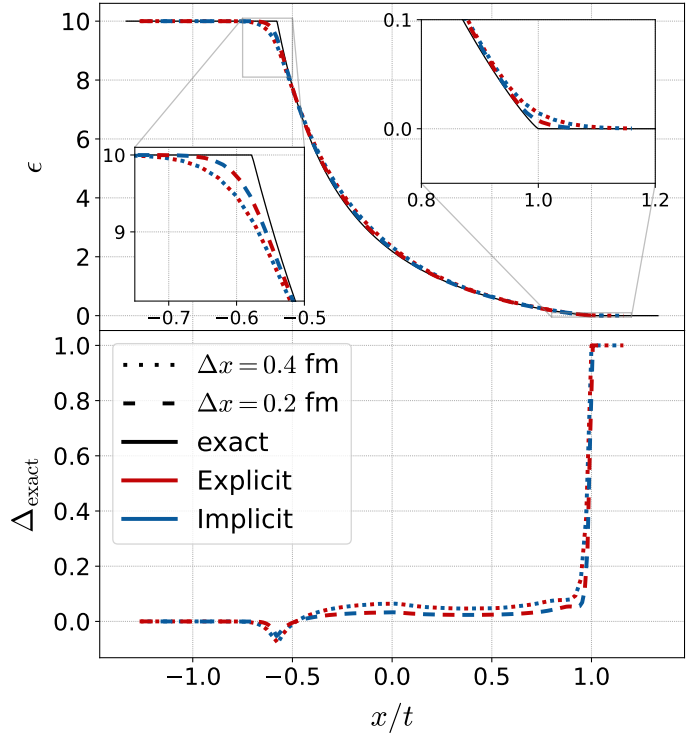}
    \caption{Same as Fig.~\ref{fig:riemannE} but $\epsilon_\text{min}=0$.}
    \label{fig:riemannVE}
\end{figure}

\subsubsection{Numerical solutions}

In the upper panel of Fig.~\ref{fig:riemannE}, we first show the numerical
result of $\epsilon(\bm{x}_j)$ at $t=t_\text{end}$ for
$\epsilon_\text{min}=1~\text{fm}^{-4}$ as a function of $x/t$. The red (blue)
curves show the results obtained by the explicit (implicit) method with $\Delta
t=0.1\times 2^{-5}\Delta x$ while the dotted and dashed lines represent the
results for $\Delta x=0.4$ and $0.2~\text{fm}$, respectively. The black solid
line shows the analytic solution $\epsilon_\text{exact}(\bm{x}_j)$. In the
lower panel, we also show $\Delta_\text{exact}(\bm{x}_j)$.

One sees that the results of the explicit and implicit methods degenerate well
at this $\Delta t$. This suggests that the error from the time discretization
is well suppressed in both methods. By comparing the results of $\Delta x=0.4$
and~$0.2~\text{fm}$, one also finds that the numerical result approaches the
analytic solution as $\Delta x$ becomes smaller. Nevertheless, a clear
deviation is still observed at the smaller $\Delta x$. This is significant
around the rarefaction front and shock at $x/t=-c_s$ and $v_\text{shock}$,
respectively, which are shown in the enlarged subpanels of the upper panel. In
particular, a large deviation $\Delta_\text{exact}\simeq0.5$ survives even for
the small $\Delta x$ at $x/t=v_\text{shock}$, which is a consequence of the
discontinuity of $\epsilon_\text{exact}(\bm{x}_j)$ at this point.

In Fig.~\ref{fig:riemannVE}, the numerical result for $\epsilon_\text{min}=0$
is shown in the same manner as Fig.~\ref{fig:riemannE}. The figure shows that
$\Delta_\text{exact}(\bm{x}_j)$ becomes almost unity in the vacuum region
$x/t>1$. As explained in Sec.~\ref{sec:error-estimate}, this result comes from
vanishing $\epsilon_\text{exact}(\bm{x}_j)$ in the vacuum region.

In the following analysis, we use $\Delta_\text{ref}(\bm{x}_j)$ defined in
Eq.~\eqref{Deltae} in place of $\Delta_\text{exact}(\bm{x}_j)$ to analyze the
time-discretization error exclusively.

\begin{figure}
    \centering
    \includegraphics[width=0.6\linewidth]{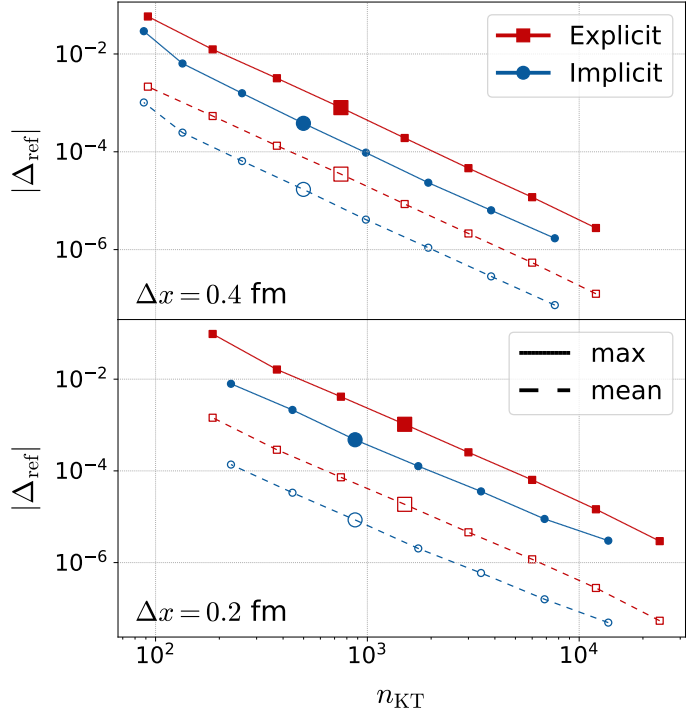}
    \caption{Maximum and average of the numerical error $|\Delta_\text{ref}|$
    defined in Eq.~\eqref{Deltae} as functions of the computational cost
    $n_\text{KT}$ for the Riemann problem with
    $\epsilon_\text{min}=1~\text{fm}^{-4}$. The maximum and average are shown
    by the solid and dashed lines, and the explicit and implicit results are
    shown by red square and blue circle symbols, respectively. The upper and
    lower panels show the results for $\Delta x=0.4$ and $0.2~\text{fm}$,
    respectively. The large points show the results at $\Delta t/\Delta x=0.1$,
    while the other small points are obtained by varying $\Delta t$ by a factor
    of $2$. Note that the absolute magnitudes of errors with different $\Delta
    x$ are not comparable because the reference $\epsilon_\text{ref}(\bm{x}_j)$
    depends on $\Delta x$.}
    \label{fig:riemannA}
\end{figure}

\begin{figure}
    \centering
    \includegraphics[width=0.6\linewidth]{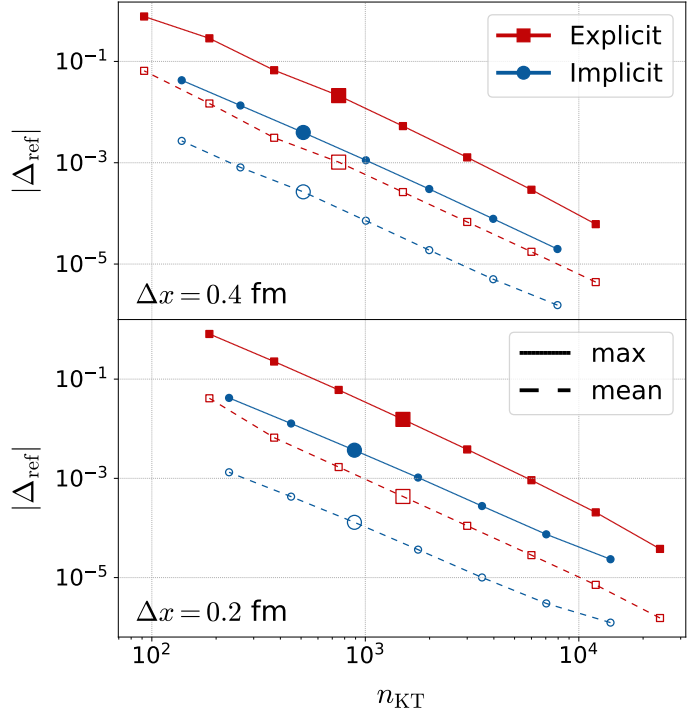}
    \caption{Same result as Fig.~\ref{fig:riemannA}, but $\epsilon_\text{min}=0$.}
    \label{fig:riemannVA}
\end{figure}

\begin{figure*}
    \centering
    \includegraphics[width=\linewidth]{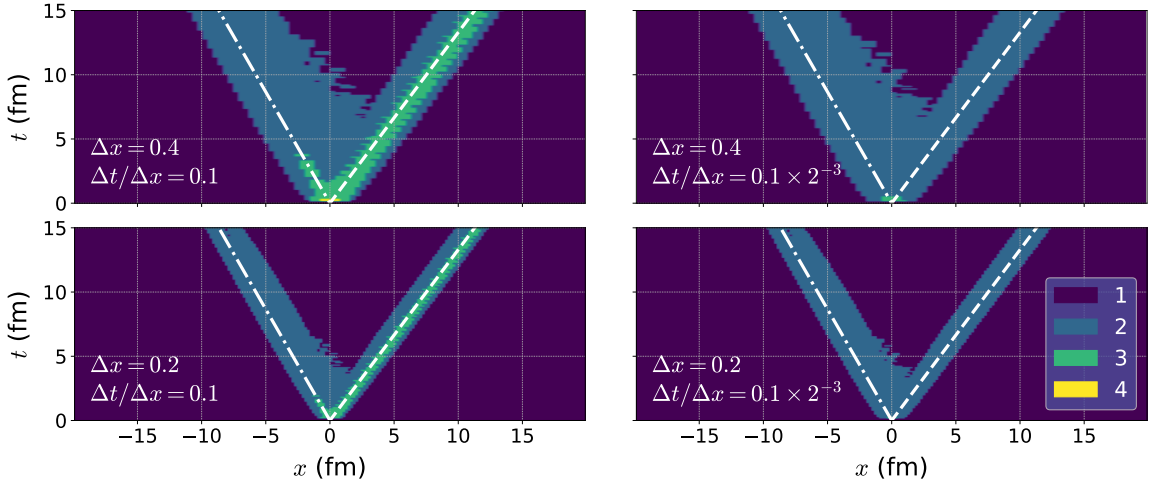}
    \caption{Number of iterations required for the fixed-point solver at each
    space-time point for the Riemann problem with
    $\epsilon_\text{min}=1~\text{fm}^{-4}$. The dashed and dash-dotted lines
    show the positions of shock and rarefaction front, respectively. Four
    results for $(\Delta t/\Delta x,\Delta x)=(0.1,0.4)$,
    $(0.1\times2^{-3},0.4)$, $(0.1,0.2)$ and $(0.1\times2^{-3},0.2)$ are shown
    in the top-left, top-right, bottom-left and bottom-right panels.}
    \label{fig:riemannC}
\end{figure*}

\begin{figure*}
    \centering
    \includegraphics[width=\linewidth]{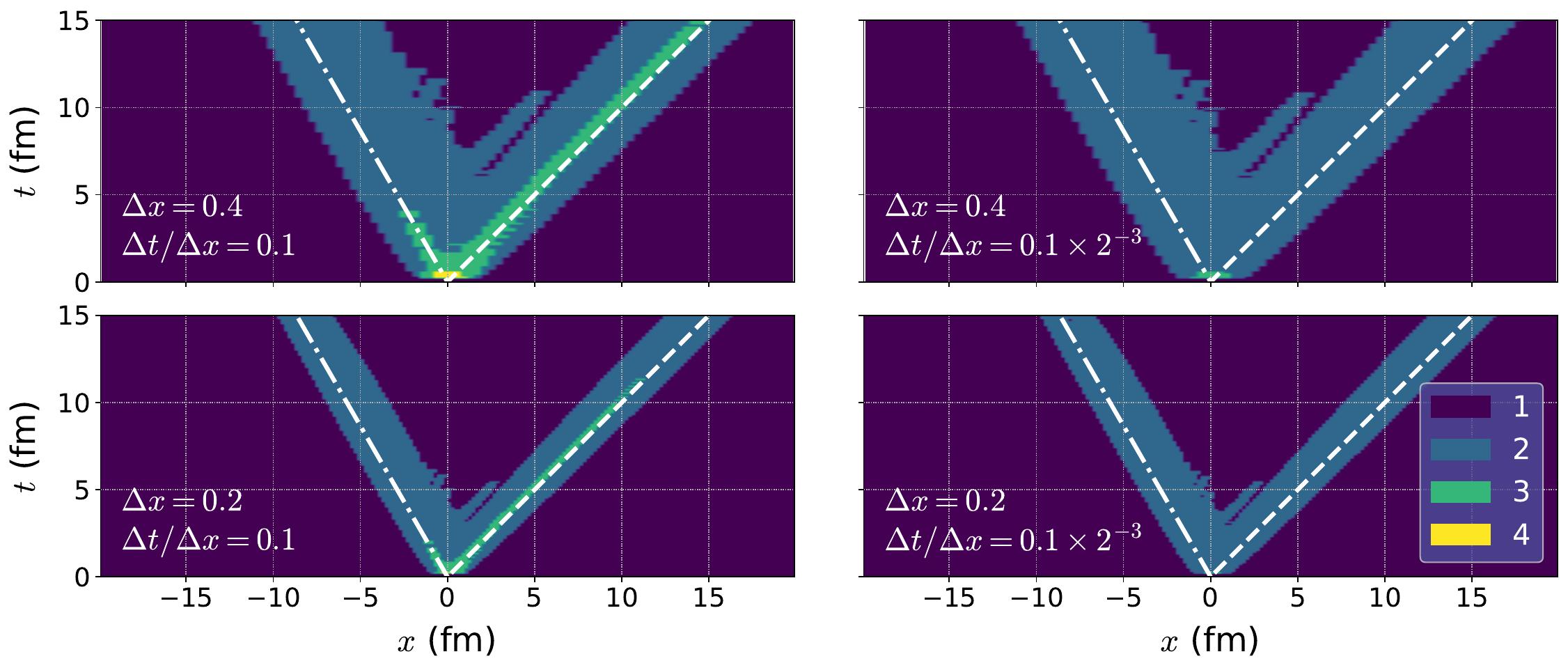}
    \caption{Same result as Fig.~\ref{fig:riemannC}, but $\epsilon_\text{min}=0$.}
    \label{fig:riemannVC}
\end{figure*}

\subsubsection{Error and computational cost}

Next, let us compare the numerical error and the cost of the explicit and
implicit methods. In Figs.~\ref{fig:riemannA} and~\ref{fig:riemannVA}, we plot
the maximum and average of the error defined by Eq.~\eqref{Deltae} at
$t=t_\text{end}$ for $\epsilon_\text{min}=1$ and $0~\text{fm}^{-4}$,
respectively, for $\Delta x=0.4~\text{fm}$ (upper plot) and $0.2~\text{fm}$
(lower plot). The horizontal axis is the computational cost $n_\text{KT}$ in
Eq.~\eqref{cost}. The red squares and blue circles show the explicit and
implicit results, respectively. The large point in each line shows the result
for $\Delta t/\Delta x=0.1$. The distance between the points along one line is
obtained from varying $\Delta t$ by a factor of two.  The numerical simulation
fails when the time step $\Delta t$ is too large, so only successful results
are shown in the figures.  In the explicit method, the cost is
$n_\text{KT}=S(t_\text{end}-t_\text{init})/\Delta t= 30~\text{fm}/\Delta t$.

We observe several notable features in Figs.~\ref{fig:riemannA}
and~\ref{fig:riemannVA}. First, all the results are approximately proportional
to $n_\text{KT}^{-2}$. This behavior is in accordance with the order of
accuracy $p=2$ of our explicit and implicit methods. Second, the computational
cost of the implicit method is significantly smaller than the explicit method
for a fixed $|\Delta_\text{ref}|$ in both maximum and average numerical errors;
The former is about $2$--$3$ times as efficient as the latter~\footnote{One
might want to compare the values of $|\Delta_\text{ref}|$ at the same $\Delta
t$ for different $\Delta x$. However, $|\Delta_\text{ref}|$ represents the
difference from $\epsilon_\text{ref}(x)$, not $\epsilon_\text{exact}(x)$, so it
does not contain the spatial discretization error. Since
$\epsilon_\text{ref}(x)$ is different for each $\Delta x$, their values at
different $\Delta x$ are not comparable.}. This means that with the same
accuracy, the implicit method is faster than the explicit one. This is a
remarkable outcome of the present study since it is widely considered that
implicit methods are slower than explicit methods.  The figures also show that
this advantage of the implicit method is obtained through the reduction both in
the numerical error and computational cost for a given $\Delta t$.

To better understand the origin of these results, in Fig.~\ref{fig:riemannC} we
show the number of iterations performed by the fixed-point solver in the
implicit method on the $t$--$x$ plane. The number of iterations is represented
by different colors for the numerical simulations with $\Delta t/\Delta x=0.1$
and $0.1\times2^{-3}~\text{fm}$ (left and right), and $\Delta x=0.4$ and
$0.2~\text{fm}$ (top and bottom). The dashed and dash-dotted lines show the
locations of the shock and the rarefaction front at which
$\epsilon_\text{exact}(x)$ and $\partial\epsilon_\text{exact}(x)/\partial x$,
respectively, have a discontinuity. The figure shows that for almost all cells
one iteration is enough to achieve the convergence. This result is reasonable
since the hydrodynamic variables would change slowly when they are smooth. As
explained in Sec.~\ref{sec:time}, we use the solution in the previous time step
for the initial guess in the fixed-point solver, and it is already a good
estimate for the next time step. Moreover, more than one iteration is required
only around the shock and rarefaction, where the number of iterations is mostly
two except for a few cells with three or four iterations. As a result, the
average number of the $[h_\text{KT}]_j$ evaluations for an update is smaller
than two in the implicit method. This makes the implicit method advantageous
compared with the explicit one where two evaluations are required for every
update.

In Figs.~\ref{fig:riemannC} and~\ref{fig:riemannVC} one finds that the regions
of two iterations have similar outlines in the left and right panels. This
behavior would be attributed to fine structures that arise due to the space
discretization.

As already discussed, in Figs.~\ref{fig:riemannA} and~\ref{fig:riemannVA} we
only show the successful results. We found that the explicit method fails for
$\Delta t/\Delta x\ge1.6$. This result is consistent with the
CFL condition~\cite{Courant1928upd, Courant1967pde}
for stability, $\Delta t/\Delta x<C_\text{CFL}$ where
$C_\text{CFL}=\mathcal{O}(1)$ depends on the detailed scheme. On the other
hand, failures of the implicit method can occur for $\Delta t/\Delta x\gtrsim0.4$. These failures happen when the
fixed-point solver does not converge. Despite $A$-stability of the implicit
method itself, the fixed-point solver is not ensured to converge for large
$\Delta t$.

\subsection{Gubser flow}

Next, we perform the same comparison for the Gubser flow as discussed in
Sec.~\ref{sec:Gubser}. We solve transverse (2+1)-dimensional hydrodynamics
assuming the boost invariance. In this case, the solution contains neither
discontinuity nor vacuum region. We employ $\epsilon_0 = 1~\text{fm}^{-4}$ and
$q = 1$ for the parameters in Eq.~\eqref{gubser} and set
$\tau_\text{init}=1~\text{fm}$ for the initial proper time. We then solve the
hydrodynamic equations numerically until the end time $\tau_\text{end} =
10~\text{fm}$.

\begin{figure*}
    \centering
    \begin{tabular}{c}
        Gubser flow, $\Delta x = 0.4~\text{fm}$\\
        \includegraphics[width=0.8\linewidth]{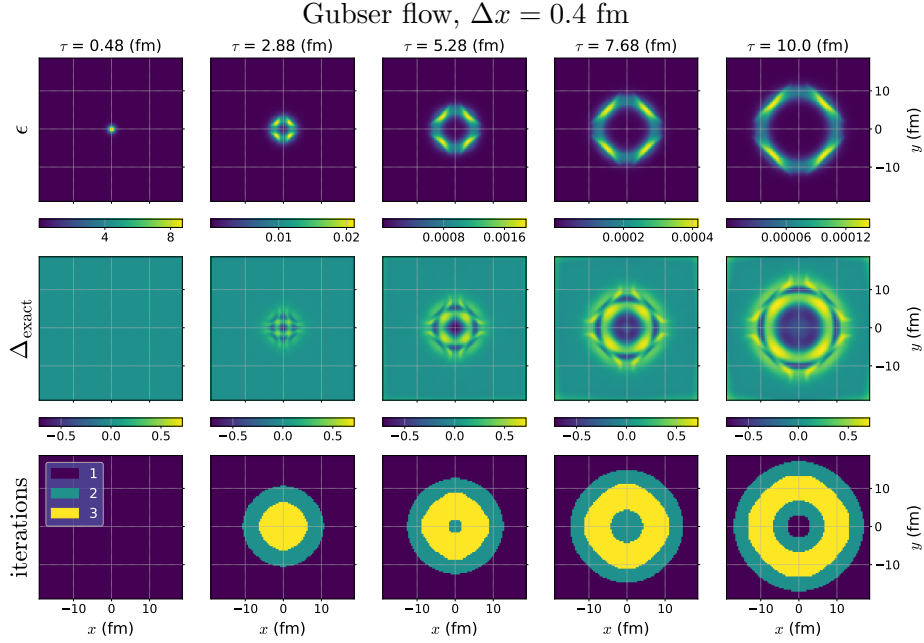} \\
    \end{tabular}
    \caption{Energy density $\epsilon(\bm{x}_j)$ (upper panels), error
    $\Delta_\text{exact}(\bm{x}_j)$ (middle panels) and the number of
    iterations of the fixed-point solver in each cell (lower panels) for the
    Gubser flow at several proper times $\tau$ for $\Delta t = 0.1\times
    2^{-5}\Delta x$ and $\Delta x = 0.4~\text{fm}$.  }
    \label{fig:gubserE2}
\end{figure*}

\begin{figure*}
    \centering
    \begin{tabular}{c}
        Gubser flow, $\Delta x = 0.2~\text{fm}$\\
        \includegraphics[width=0.8\linewidth]{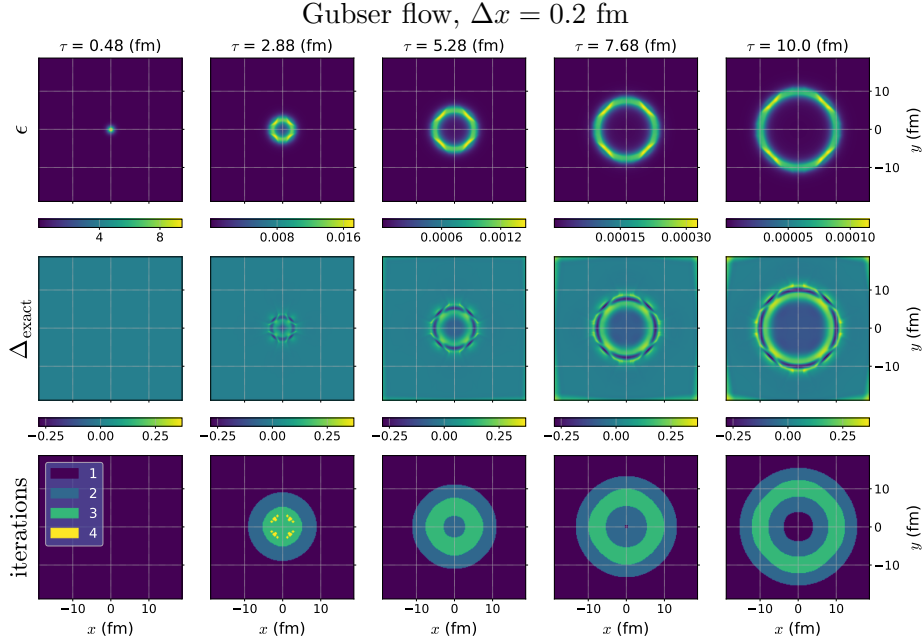} \\
    \end{tabular}
    \caption{Same results as Fig.~\ref{fig:gubserE2}, but with $\Delta x = 0.2~\text{fm}$.}
    \label{fig:gubserE1}
\end{figure*}

In the top panels of Figs.~\ref{fig:gubserE2} and~\ref{fig:gubserE1}, we first
show the energy density at several values of $\tau$ for $\Delta x=0.4$ and
$0.2~\text{fm}$, respectively, obtained by the implicit method with $\Delta
t=0.1\times 2^{-5}\Delta x~\text{fm}$.  Here we do not show the result of the
explicit method since its behavior is almost the same as anticipated from the
discussion in Sec.~\ref{sec:result:Riemann}. The panels show that the numerical
solution has non-circular behaviors, while the exact solution,
Eq.~\eqref{gubser}, has azimuthal symmetry. Correspondingly,
$\Delta_\text{exact}(\bm{x}_j)$ shown in the middle panels has oscillating
behaviors. These errors are attributed to the space discretization, as we have
checked that their structure hardly changes for small $\Delta t$. By comparing
Figs.~\ref{fig:gubserE2} and~\ref{fig:gubserE1}, one also finds that the
deviation from the exact solution is suppressed as $\Delta x$ becomes smaller.

Shown in the lower panels of Figs.~\ref{fig:gubserE2} and~\ref{fig:gubserE1}
are the number of iterations in the fixed-point solver for each $\tau$. These
panels show that two iterations are needed around the region where
$\epsilon(\bm{x}_j)$ changes rapidly, but only one iteration is enough for the
other wide area. This result is consistent with the one in
Sec.~\ref{sec:result:Riemann}.

\begin{figure}
    \centering
    \includegraphics[width=0.6\linewidth]{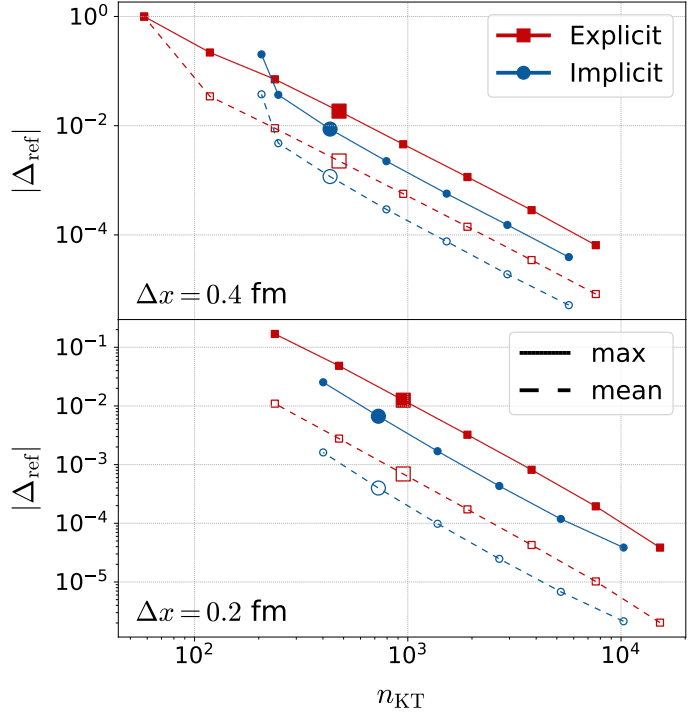}
    \caption{Same as Fig.~\ref{fig:riemannA}, but for the analysis of the Gubser flow in Figs.~\ref{fig:gubserE2} and~\ref{fig:gubserE1}.}
    \label{fig:gubserA}
\end{figure}

In Fig.~\ref{fig:gubserA}, we compare the maximum and the average of
$|\Delta_\text{ref}(\bm{x}_j)|$ in the explicit and implicit methods as
functions of $n_\text{KT}$. The meanings of the lines and symbols are the same
as before. The figure shows that the implicit method is advantageous again in
reducing the computational cost even in this case.

\subsection{Event-by-event initial condition for heavy-ion collisions}

Finally, we apply the same analysis to realistic cases of the relativistic
heavy-ion collisions by generating initial conditions using
\TRENTo{}~\cite{Trento}.  We perform the (2+1)-dimensional event-by-event
simulations for lead-lead collisions with impact parameter $b=3~\text{fm}$ and
collision energy $\sqrt{s_\text{NN}} = 2.76~\text{TeV}$\@. We use
$\tau_\text{init}=0.48~\text{fm}$ as the initial time of the hydrodynamics. For
the other parameters of \TRENTo{}, we take the same values as in
Refs.~\cite{Bayesian,BayesianNature}.  For the EOS, We use the parametrized
form~\cite{Bazow:2016yra, Bazow:github-gpu-vh} of the Wuppertal-Budapest
lattice EOS~\cite{Borsanyi:2010cj}.  We then solve the hydrodynamic evolution
until $\tau_\text{end} = 10~\text{fm}$.

\begin{figure*}
    \centering
    \begin{tabular}{c}
        \TRENTo{}, $\Delta x = 0.4~\text{fm}$\\
        \includegraphics[width=0.8\linewidth]{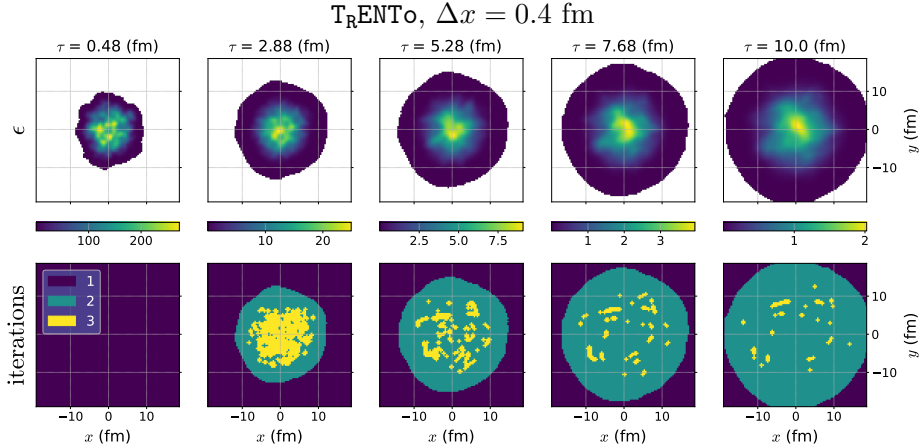} \\
    \end{tabular}
    \caption{Energy density $\epsilon(\bm{x}_j)$ (upper panels) and the number
    of iterations of the fixed-point solver in each cell (bottom panels) for a
    \TRENTo{} initial condition (lead-lead collision with $b=3~\text{fm}$ and
    $\sqrt{s_\text{NN}} = 2.76~\text{TeV}$) at several values of proper time
    $\tau$ for $\Delta t = 0.1\times 2^{-5}\Delta x$ and $\Delta x =
    0.4~\text{fm}$.}
    \label{fig:trentoE2}
\end{figure*}
\begin{figure*}
    \centering
    \begin{tabular}{c}
        \TRENTo{}, $\Delta x = 0.2~\text{fm}$\\
        \includegraphics[width=0.8\linewidth]{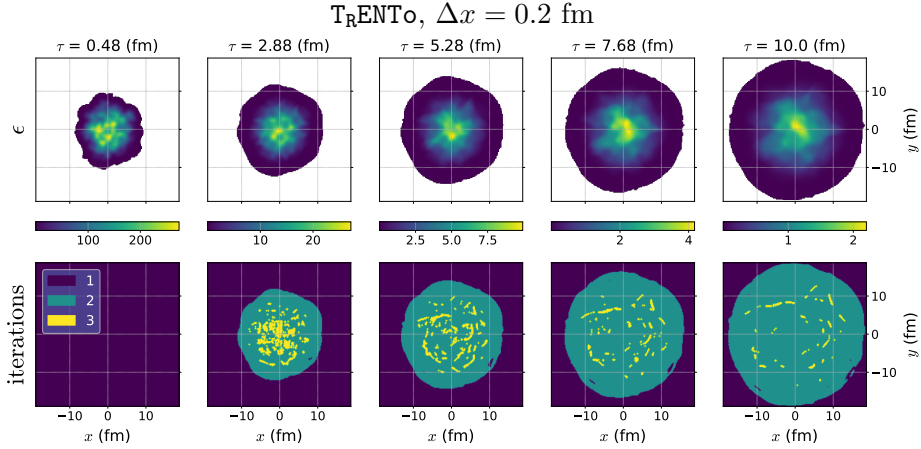} \\
    \end{tabular}
    \caption{Same as Fig.~\ref{fig:trentoE2}, but for $\Delta x = 0.2~\text{fm}$.}
    \label{fig:trentoE1}
\end{figure*}

In the upper panels of Figs.~\ref{fig:trentoE2} and~\ref{fig:trentoE1}, we show
the transverse profile of energy density $\epsilon(\bm{x}_j)$ for one event in
the $x$--$y$ plane obtained with the finest $\Delta t = 0.1\times 2^{-5}\Delta
x$ in the implicit method for $\Delta x=0.4$ and $0.2~\text{fm}$,
respectively. The cells satisfying $\epsilon(\bm{x}_j)<\epsilon_\text{thr}$ are
shown by the white color. As discussed in Sec.~\ref{sec:error-estimate}, these
cells are discarded in the error analysis. In the lower panels, the
corresponding number of iterations in the fixed-point solver is shown. These
results again show that the average number of iterations is smaller than two,
which is a reasonable result in light of the arguments in the previous
subsections.

\begin{figure}
    \centering
    \includegraphics[width=0.6\linewidth]{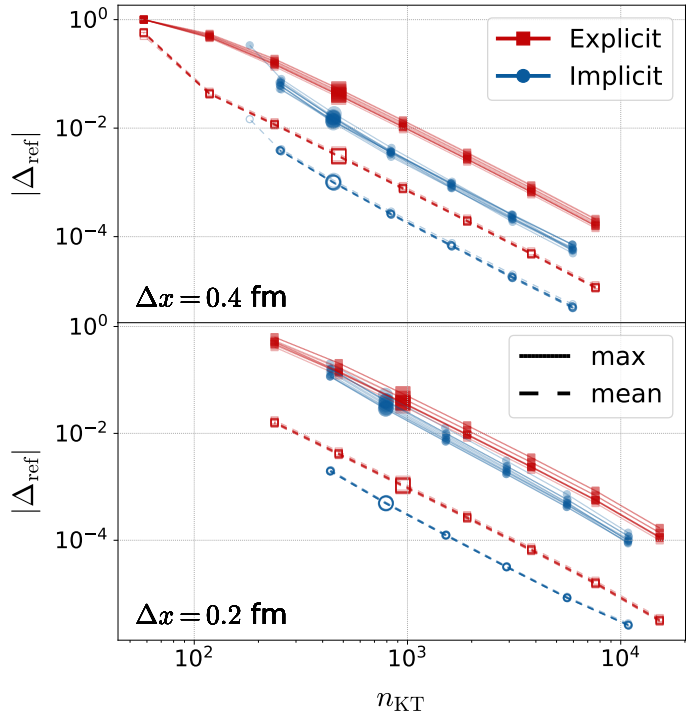}
    \caption{Same as Fig.~\ref{fig:riemannA}, but for $10$ initial conditions generated by \TRENTo{}.}
    \label{fig:trentoA}
\end{figure}

In Fig.~\ref{fig:trentoA}, we plot the maximum and the average of
$|\Delta_\text{ref}|$ as functions of $n_\text{KT}$ for ten initial conditions
generated randomly by \TRENTo. The figure shows that the implicit method tends
to carry out the analysis with a smaller computational cost than the explicit
method for a given error, although its advantage is less significant compared
with the results in previous sections.

\section{Conclusion}

In this study we proposed a new method that numerically solves relativistic
hydrodynamics with an implicit Runge-Kutta (RK) time integrator and implemented
an efficient solver for ideal hydrodynamics. The one-stage Gauss-Legendre
method is employed as the implicit method while the Kurgamov-Tadmor (KT) scheme
is used for the space discretization. To solve the implicit RK equations, we
propose a simple fixed-point solver with local optimization. For the initial
guess of the solver, we employ the solution in the previous RK step. We also
showed that our locally optimized implicit method can be viewed as an adaptive
mixing of two second-order methods: it corresponds to the two-step
Adams-Bashforth method when a single iteration is performed in smooth regions,
while it approaches the one-stage Gauss-Legendre method as the number of
iterations increases in stiffer regions. As a result, it benefits from the
efficiency of the two-step Adams-Bashforth method and the stability of the
one-stage Gauss-Legendre method.

The performance of our method was compared with Heun's method, which is one of
the conventional explicit methods, with the same space discretization for
different initial conditions in one and two space dimensions. We found that the
computational cost of the implicit method is always smaller than the explicit
one to attain the same accuracy except for impractically large $\Delta t/\Delta
x$. The advantage of the former mainly comes from the fact that the fixed-point
solver requires less than two iterations on average for convergence. In
comparison, Heun's method as a two-stage RK method needs two substeps in each
RK update. This makes the number of evaluations of the KT operator smaller in
the implicit case.  These results show that our implicit method can be
efficient and useful for the application to HIC\@.

This also suggests that the implicit RK method can be used for more general
setups of computational hydrodynamics to attain efficient and simple
implementations.  Because implicit methods generally have an advantage in their
numerical stability, they should lead to better implementation of the
relativistic hydrodynamics with less technical procedures for stabilization,
such as the limiters and regulations of variables.  This advantage will be more
effective in dealing with more complex equations such as fluctuating
hydrodynamics and magnetohydrodynamics.

Although we limited our analysis to ideal hydrodynamics throughout this paper,
it is important to extend the analysis to (3+1)-dimensional viscous
hydrodynamics and to incorporate conserved charges for practical applications
to HIC\@.  Our implicit method has a problem
in the convergence of the fixed-point method at large $\Delta t/\Delta x$.  A
solution to improve the convergence will be discussed in a future
publication.  We have already started an implementation of viscous
hydrodynamics in (3+1) dimensions with the implicit method. As will be reported
in the forthcoming publication, our preliminary analyses show that our method
is advantageous even in this case.  We also confirmed the efficiency of our
method in solving the diffusion equation and the Klein--Gordon equation.  These results
indicate the advantage of our method in a broader class of problems.

\section{Acknowledgments}
The authors thank Marcus Bluhm and Gregoire Pihan for fruitful discussions in
the initial phase of this work.  N.~T. is supported by MEXT (Ministry of
Education, Culture, Sports, Science and Technology). This work was supported by
JSPS KAKENHI (Grant Nos.~JP19H05598, JP20H01903, JP22K03619, JP23H04507, and
JP23K13102) and the TYL-FJPPL program of IN2P3-CNRS and KEK. M.~Nahrgang and
N.~Touroux acknowledge the support of the program ``Etoiles montantes en Pays
de la Loire 2017''.

\appendix

\section{Comparison with the midpoint method}
\label{sec:midpoint}

\begin{figure}
    \centering
    \includegraphics[width=0.6\linewidth]{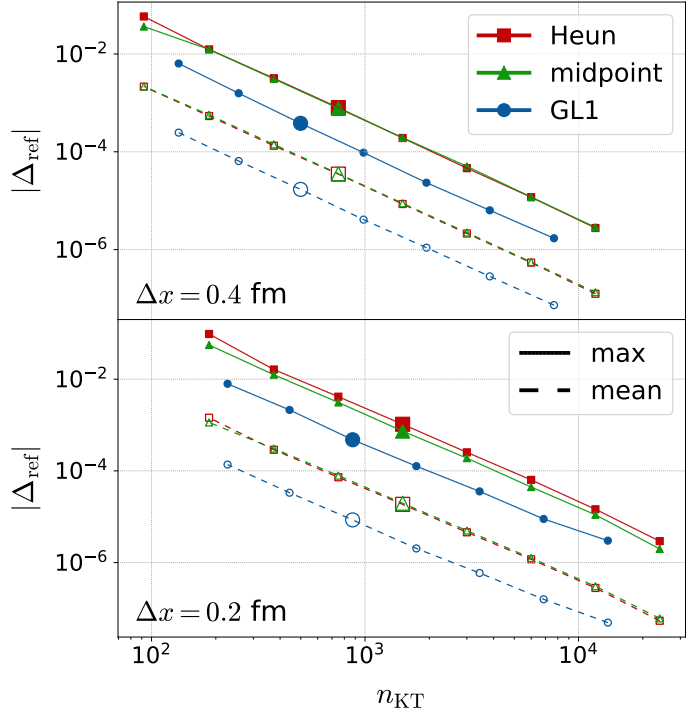}
    \caption{Maximum and average of the numerical error $|\Delta_\text{ref}|$
    as functions of $n_\text{KT}$ for the Riemann problem with $\emin =
    1~\text{fm}^{-4}$. The result of the midpoint method is shown by the green
    triangle symbols. The results for Heun's and GL1 methods are the same as
    Fig.~\ref{fig:riemannA}.}
    \label{fig:midpoint1}
\end{figure}
\begin{figure}
    \centering
    \includegraphics[width=0.6\linewidth]{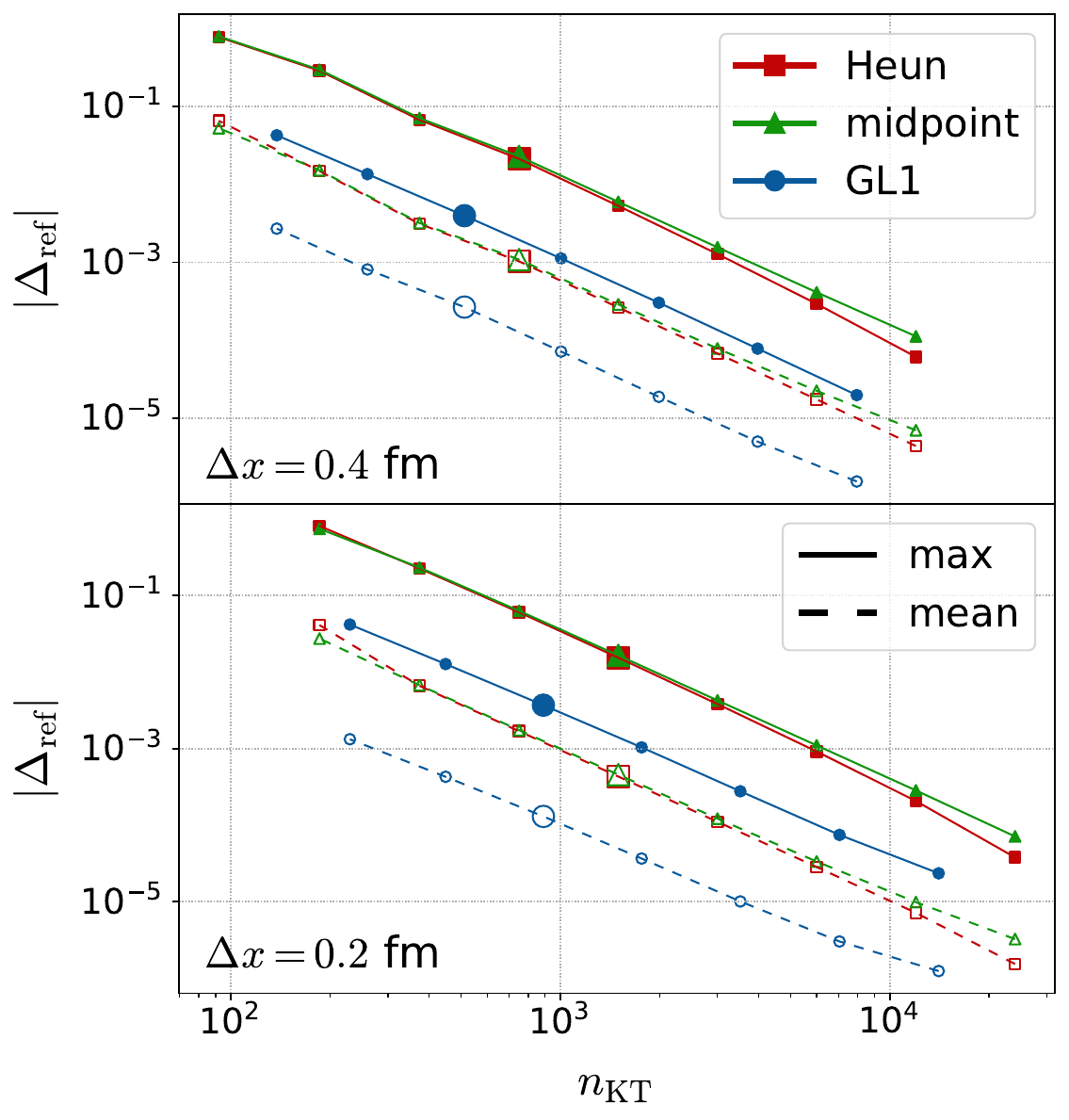}
    \caption{Same result as Fig.~\ref{fig:midpoint1}, but $\emin = 0~\text{fm}^{-4}$.}
    \label{fig:midpoint2}
\end{figure}

In Sec.~\ref{sec:results}, we compare the computational efficiencies of Heun's
and GL1 methods as representatives of explicit and implicit methods,
respectively. In this appendix, we employ the midpoint method as another
explicit method and compare its efficiency with Heun's and GL1 methods. The
Butcher table of the midpoint method is shown in the middle column of
Table~\ref{tab:butcher}. As shown there, this method has the same order of
accuracy $p=2$ as Heun's and GL1 methods.

In Figs.~\ref{fig:midpoint1} and~\ref{fig:midpoint2}, we plot the maximum and
the average of the numerical error $|\Delta_\text{ref}|$ as functions of
$n_\text{KT}$ for the (1+1)-dimensional Riemann problem with
$\epsilon_\text{min}=1$ and $0~\text{fm}^{-4}$, respectively. The results show
that the performance of the midpoint method is almost the same as Heun's
method.

\bibliographystyle{elsarticle-num}
\bibliography{biblio.bib}

\end{document}